\documentstyle[aps,pre,twocolumn,epsfig]{revtex}

\begin{document}

\title{Chemical or biological
activity in open chaotic flows}
\author{Gy\"orgy K\'arolyi$^{1}$,
\'Aron P\'entek$^{2}$,
Zolt\'an Toroczkai$^{3,4}$,
Tam\'as T\'el$^{4}$, and
Celso Grebogi$^{5}$
}

\address{
$\mbox{ }^1$Department of Civil Engineering Mechanics,
Technical University of Budapest, M\H{u}egyetem rkp.~3,
H-1521 Budapest, Hungary \\
$\mbox{ }^2$Institute for Pure and Applied Physical Sciences,
University of California
at San Diego, La Jolla, CA 92093-0360, USA \\
$\mbox{ }^3$Center for Stochastic
Processes
in Science and Engineering,
and Department of Physics,
Virginia Polytechnic Institute, 
Blacksburg, VA 24061-0435, USA\\
$\mbox{ }^4$Institute for Theoretical Physics,
 E\"otv\"os University, Puskin
utca 5-7, H-1088 Budapest, Hungary \\
$\mbox{ }^5$
Institute for Plasma Research, University of Maryland,
College Park, MD 20742, USA}

\date{\today}

\author{
\begin{center} \small{
\parbox[t]{13.5cm}{
\qquad We investigate the evolution
of particle ensembles in open 
chaotic hydrodynamical flows. Active processes
of the type $A+B \to 2B$ and $A+B
\to 2C$ are considered in the limit of weak diffusion.
As an illustrative
advection dynamics we consider a
model of the von K\'arm\'an vortex street, a time
periodic two-dimensional flow of a viscous fluid around
a cylinder. We show that a fractal unstable
manifold acts as a catalyst for the process,
and the products cover fattened-up copies of this manifold.
This may account for the observed filamental intensification of activity
in environmental flows.
The reaction equations valid in the wake
are derived either in the form of dissipative maps
or differential equations depending on the regime under consideration.
They contain terms that are not present in the traditional
reaction equations of the same active process: the decay of
the products is slower while the productivity is much faster than
in homogeneous flows.
Both effects appear as a consequence of underlying
fractal structures.  
In the long time limit, the system locks itself
in a dynamic equilibrium state synchronized to the flow
for both  types of reactions.
For particles of finite size
an emptying transition
might also occur leading to no products  left in the wake.
}}
\end{center}
}

\maketitle



\newpage
\section{Introduction}

Active processes taking place in
chaotic hydrodynamical flows
have attracted recent interest
\cite{Metcalfe}-\cite{Neufeld}.
By chaotic we mean time-dependent but
nonturbulent velocity fields with chaotic tracer dynamics
(Lagrangian chaos) \cite{Ottino}-\cite{CSF}.
In the simplest approximation we can assume that
the advected particles undergo 
certain chemical or biological changes
but do not modify the fluid flow.
The motivation for such studies has been to understand
the effects of { imperfect mixing} \cite{Epstein}
due to the underlying
chaotic particle dynamics. The implications can be perceived
in laboratory experiments \cite{Epstein,Kondep}, but the
effects are perhaps  more striking in
environmental flows.
In particular, there is increasing evidence of filamental
structures in the product 
distribution of environmental processes
both in the atmosphere, such as ozone reactions
\cite{Legras,Chipper,Haynes,Tan,Mariotti}, and 
resulting from the interaction of  populations of
microorganisms in the sea, such as plankton distributions
\cite{Holligan,Gibson}. Our aim is to
show that these structures might be consequences
of the fractal structures of the reaction free flows
\cite{TKPTG}.

Here we shall consider open flows with asymptotic simplicity
in which the velocity field in the far up and downstream regions
is uniform. A well-known (time-periodic) laboratory example
is the flow around
a cylinder. Its actual realization  can be observed in
environmental flows, like, e.g., in the fluid motion in the
wake of a pillar or in the motion of air behind an isolated mountain.
A unique feature of such open flows
is the pronounced and stable fractal feature associated with
the chaotic tracer dynamics \cite{ROMKED}-\cite{PA}, which is
clearly measurable in
experiments \cite{Sommerer}.
The central object governing the dynamics is a nonattracting chaotic
saddle  \cite{T0} containing an infinite number of periodic
and nonperiodic tracer orbits
which remain bounded and never reach either the far up or the
downstream regions.
A characteristic quantity of the saddle is its escape rate
$\kappa$ whose reciprocal value is the average chaotic lifetime.
The far up and downstream regions are
foliated by the saddle's stable and
unstable manifold, respectively.
The saddle's
unstable manifold directs tracers  ever approaching the
saddle to the far downstream region.
Though both the saddle and its manifolds are
not space-filling fractal objects,
it has been pointed out \cite{ROMKED} - \cite{Sommerer}
that the unstable manifold
is the avenue of propagation and transport in such flows.
It is the pronounced fractal structure of such flows
(which is not present in closed flows) that makes
them specially interesting catalyst of active processes
\cite{TKPTG}.

In this paper we consider the advection of {\em active} particles
in flows with asymptotic simplicity in which the 
activity is assumed to be of chemical or biological origin
in the simplest possible form. The reaction
is a kind of ``infection''  leading to a
change of certain properties, such as color  of reacting particles.
Particles with new properties are the {\em products}.
Since it is in the close vicinity of the chaotic saddle 
and its unstable manifold that the particles spend the longest time
close to each other, it is there where the effect of the activity
is most pronounced. It is then natural
to expect that the  { products}
should accumulate along the unstable manifold and trace out this
fractal object.

In our work  we support this conjecture
and present a detailed analysis of such
active processes. We show that the unstable manifold
of the chaotic saddle is the {\em backbone} of the reaction.
The newly born components
{ cover the branches of the unstable manifold}
with a well defined  {\em average width}
$\varepsilon^*$. Thus, an
 effective {\em fattening up}
of the fractal set takes place due to the activity of the tracers.
This implies that on linear scales smaller
than this width $\varepsilon^*$, fractality is washed out, but
a clear fractal scaling of the material with a dimension $D_0$
can be observed
on larger scales.
This fractal dimension is the {\em same}
$D_0$ as that of the unstable manifold in the reaction
free flow.
Although the fractal set itself is a set of measure zero,
the amount of chemical products
is {\em finite}  due to the fattening-up process
of this manifold.

A consequence of the fractal backbone is that
the amount of the reaction product follows a singular scaling law with
{\em irrational} $D_0$-dependent powers of the number of product
particles, signaling
a {\em singular} enhancement of productivity \cite{TKPTG}.
(The enhancement of activity is meant in comparison with non-chaotic,
e.g., stationary flows.) 
This singularly enhanced rate of activity has profound practical
consequences. It may account for the observed filamental patterns
of intense activity
in environmental flows \cite{Legras,Chipper,Haynes,Tan}, an effect 
that cannot be explained if
one considers diffusion processes alone.
In this work we show how small-scale structures are generated in
the dynamics of active particles, and how these dynamical structures
are responsible for the enhancement of activity.

In summary, the effect of the chaotic saddle producing this activity
is twofold: (i) to keep the reacting particles longer,
as given by the escape rate $\kappa$,
in the interaction region, and (ii) to generate the unstable
manifold's fractal structure that brings the reacting particles together.

We derive the corresponding {\em reaction equations}
in the form of maps or differential equations depending on the 
regime under consideration.
Such processes are generalization of classical surface
reactions \cite{Landau},
but, by contrast, in our case the surface is a fattened-up fractal.
The reaction equations contain new terms {\em not}
present in the traditional well-stirred reaction model
of the same process.
In spite of the passive tracers'
Hamiltonian dynamics, these reaction equations turn out to be of
{\em dissipative} character possessing  attractors. 

We find that the chemical 
activity and the advection by the hydrodynamical flow
are in permanent {\em competition}. Due to this competition,
most typically,
a kind of {\em stationary state} sets in after
sufficiently long times. 
In the case of time-periodic flows
of period $T$, the
asymptotic state is typically also periodic with $T$, i.e.,
the reaction becomes {\em synchronized to the flow}.
Thus, a chaotic particle dynamics
is consistent with a non-chaotic reaction dynamics.

To be more specific, we consider simple {\em kinetic}
models \cite{Metcalfe} with disk-like particles. 
Two particles of different kinds undergo a reaction
if and only if they come within a distance $\sigma$, which 
is the interaction range. Due to the incompressibility
of the fluid (which is always a good approximation 
for velocities much below the speed of sound),	
two-dimensional flows are {\em area-preserving}.
Therefore, if we cover a layer of fluid by particles, this single-particle
coverage of the layer will not change in time.
In other words, the average linear distance $\varepsilon_0 \le \sigma$
between the actual nearest-neighbor particles is constant. 
(Note, however, that the linear distance between specific particles is
typically increases in time due to the chaotic dynamics.)
Thus, the result of the reaction is
just the spreading of the products in the layer of particles being
advected by the flow.
We emphasize again that particles are assumed to
have no feedback on the flow. Furthermore,
the advection dynamics is purely deterministic, i.e., we work in the 
limit of weak  diffusion where the reaction range represents
some kind of diffusion distance too.

We shall consider both an autocatalytic process,
$A+B \rightarrow 2 B$, and
a collisional reaction $A+B \rightarrow 2 C$. In both cases
$A$ is considered to be the background material which covers initially
the full infinite layer of observation.
In the autocatalytic process a single seed
of particle $B$ is sufficient to trigger  reactions
leading to the stationary state, while in the collisional
reaction a continuous feeding  of material $B$
is necessary.

For computational convenience
we assume that the reactions are instantaneous and
take place at integer multiples of a time lag $\tau$.
We shall see that an important dimensionless parameter will be
the ratio between the time lag and the average chaotic lifetime:
$\nu=\tau/(1/\kappa)=\tau \kappa$. This can also be considered to be
the dimensionless reaction time, whose reciprocal value
tells us how many reaction events occur
on the characteristic time of chaos.

The case of time continuous reactions is obtained in
the limit $\tau \rightarrow 0$ (or
$\nu \rightarrow 0$) by keeping the {\em reaction front velocity} $v_r$
finite.
In this limit we assume, too, that the average distance
$\varepsilon_0$ between particles goes also to zero,
and we obtain a continuous distribution of particles.
We call this limit the {\em chemical frame}.
A fractal product distribution is then expected to appear if
the reaction is {\em slow} compared to the flow ($v_r\ll \kappa L$ with
$L$ as a characteristic length). 
An example for time continuous reactions is related to
the depletion of ozone at 
the polar vortex: the trimolecular reaction of ClO with
NO$_2$.  In late winter and early spring the polar vortex exhibits high
concentrations of ClO and very low concentrations of NO$_2$ while outside the
vortex the situation is typically reversed with relatively high concentrations
of NO$_2$ and low concentrations of ClO.
Thus the reaction ClO$+$NO$_2
\rightarrow$ ClONO$_2$ is a natural candidate to produce a filamental ClONO$_2$
distribution along the edge of the polar vortex
\cite{Chipper,Tan,Mariotti,McKenna,Mariott}
on the time
scale of a few days, where the molecular diffusivity is negligible.

\begin{figure}[htbp] 
\centering{\epsfig{file=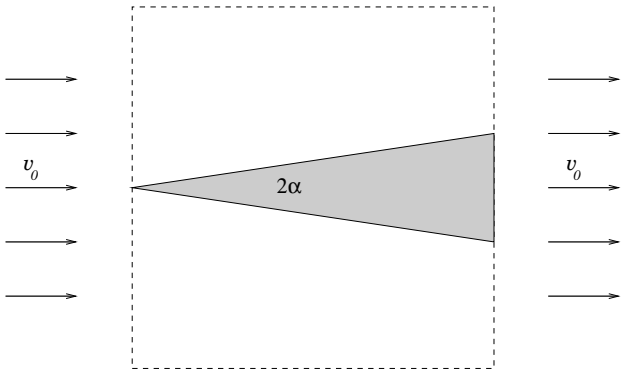,width=0.95\hsize}}
\vspace*{0.5cm}
\caption{Schematic diagram of a uniform flow
of velocity $v_0$  with autocatalytic reaction.
A  single seed $B$ is kept fixed at the left corner of the gray triangle
occupied by material $B$
which lies in a layer of white
$A$ particles. The observation region is a unit square
(dashed line).}\label{F:Schem}     
\end{figure}

In order to contrast our work with the 
conventional description of chemical processes, we briefly
discuss now the continuous time dynamics of
the autocatalytic surface reaction
$A+B \rightarrow 2 B$ in a uniform flow. Let us observe
the flow moving to the right with velocity $v_0$ in
a unit square (fixed to the observer at rest). A seed particle of type
$B$ is kept fixed about the middle of the left boundary. Particles
of type $A$ are distributed with uniform density everywhere on the surface
of the flow (also upstream). The seed particle starts to interact with
its $A$ neighbors transforming them into $B$. Since
$B$ particles are transported away, more and more 
$A$ particles are converted into $B$. Let us assume that
at time $t$ the $B$ particles cover a triangle across
the square which is symmetric about
a horizontal line (Fig.~\ref{F:Schem}).

We assume that the half angle $\alpha$ is small. The area
${\cal A}_B$ occupied by $B$ is simply $ \alpha$.
The change in this area during time $dt$ is due to a horizontal
displacement $v_0 dt$ and a vertical increase $v_r dt$
of both fronts, where $v_r$
is the {reaction front velocity}.
The gain of the area
${\cal A}_B$ is just
$2 (-v_0 \alpha +v_r) dt=$
$2 (-v_0 {\cal A}_B + v_r) dt$. The differential equation
governing this area is thus
\begin{equation}
\dot{{\cal A}}_B=-2 v_0 {\cal A}_B+2 v_r.
 \label{diffeqa}
\end{equation}
It has a stationary solution, corresponding to a stabilized
triangular distribution of $B$ particles  of area
${\cal A}_B^* \equiv \alpha^*=v_r/v_0$.
We shall see that the presence of a  saddle in a time dependent nonuniform
flow results in a slower decay and a faster production
($2 v_0$ is replaced by the escape rate
$\kappa$ and the production term will contain a factor with a negative power
of the area itself due to the fractality of the
unstable manifold).

For generality, we also investigate
cases where the time lag $\tau$ is finite so that its dimensionless
version
$\nu$ is of order unity or larger.
If $\nu$ exceeds a critical value, 
we find that 
no product remains, 
i.e., an {\em emptying transition} takes place.
Because of the finite values 
of $\tau$, and the discrete
character of the particles
($\varepsilon_0$ might also be considered as the size of particles),
this latter effect might be of relevance to
biological processes accompanying advection.
 An example can be a crude 
model of the dynamics of
plankton populations
\cite{Holligan,Gibson} in the presence of
a time dependent flow.
The so-called zooplanktons ($B$) have a daily
rhythm: they sink down  during night 
time  but come up to the surface
of the sea
again during day-time when they  
eat up phytoplanktons ($A$),
reproduce themselves and then grow  in number.

This paper is organized as follows. In
Sec.\ II, we present
the model and  the numerical procedure
used, while the results
for both reaction types are shown in  Sec.\ III.\@ A detailed theory 
based on these observations  is
derived in Sec.\ IV.\@ 
The concluding Sec.\ V\@ gives remarks on  
properties expected to be
valid in more general models of 
active processes in open flows.


\section{The model flow and numerical procedure}

The flow chosen to illustrate the 
fractal active dynamics is an example
of a two-dimensional, incompressible time-periodic fluid motion,
the case of the von K\'arm\'an
vortex street in the wake of a cylinder \cite{SPO}-
\cite{Sommerer}.
The radius $R$ of the cylinder and 
the period $T$ of the flow
are taken as the length unit and the 
time unit, respectively.
In what follows we keep the flow 
parameters constant, implying a fixed
value of the escape rate $\kappa$, 
and investigate the dependence of the
reaction outcome on parameters 
like the reaction range ($\sigma$)
and  time $(\nu)$.

For simplicity, we use an analytic model for the stream
function introduced in Ref.~\cite{JTZa}
(the explicit form of the stream function
can also be found in the Sec. III of Ref. \cite{PTTGY}).
This
model has been motivated
by direct numerical simulations at 
Reynolds number of about 250 \cite{JZ}, and has been
used successfully to reproduce 
qualitative features of the
tracer dynamics. The escape rate of the  
particles in the reaction free
flow is $\kappa=0.36$ and the fractal 
dimension of the unstable manifold
is $D_0=1.61$, while the background 
flow velocity is $v_0=14$
\cite{JTZa,PTTGY}.

Since the flow is periodic, we 
fix the ``phase'' of the reaction
relative to the flow. We 
consider time zero, $t=0$, to
be the instant when a vortex is 
 born close to the
surface of the first quadrant of the
 cylinder and, simultaneously,
a fully developed vortex is detaching 
in the fourth quadrant of the surface \cite{JTZa,PTTGY}.

For convenience,
 we carry out the simulations 
on a uniform 
rectangular grid of lattice size
$\varepsilon_0$ covering both the
incoming flow and the mixing region
in the wake of the cylinder. This $\varepsilon_0$
also corresponds to the average distance between nearest-neighbor particles.
If there is a tracer inside a cell, it is
always considered to be in its 
center; the reaction range
is bounded from below by the 
lattice size: $\varepsilon_0 \le \sigma$.
This projection of
the tracer dynamics on a grid essentially 
defines a mapping 
among the
cells. 

The course of the chemical reaction
starts with nearly all cells occupied
by  species $A$, the background material.
Few cells contain $B$ distributed 
according to the initial conditions chosen for the type of reaction
under consideration.
One iteration of the process just described consists of two mappings
in involution. The first mapping models the advection
of the particles on the chosen grid, while the
second models the instantaneous active process (e.g. chemical reaction)
occurring on the same grid of cells.
In fact, in any closed region considered there is a loss 
of the products due to the advection but also a
gain in the product amount due to the reaction.
Later we find a balance between these two competing
effects manifested in a steady state in the product distribution.
The simulation consists of a 
repeated application of advection and reaction steps.
We apply different algorithms for different reaction types.

\subsection{Autocatalytic reaction: $A+B \rightarrow 2 B$}

If a tracer starting from the center of a cell
is advected into another one after time $\tau$, then
the latter cell is considered to be the image
of the first one with respect to the dynamics.
After an application of the map, a cell will be
considered occupied by reagent $B$
if it is an image of at least 
one  $B$ cell. Otherwise the cell
is considered to contain species 
$A$ after the mapping. In addition,
if a cell contains $B$ at the time of the reaction,
all of the 8 neighboring cells are infected by $B$.
Consequently,
$ \varepsilon_0$  plays the 
role of the interaction range $\sigma$
in our simulation.

\subsection{Collisional 
reaction: $A+B \rightarrow 2 C$}

In this case, a cell is considered to be the image 
of another one with respect to the dynamics
if its center's preimage
is inside this other cell $\tau$ time earlier \cite{note}.
This defines the mapping among the cells due to the dynamics.
After the action of the mapping, the reaction can modify the cell contents:
Any cell containing $A$ ($B$) before the reaction
becomes $C$, if there is a $B$ ($A$) cell within  a radius $\sigma$ from its
center.
Otherwise the cell keeps its content.
Numerically we found convenient to store
the configuration of the lattice 
just before reactions only. Then the content of a cell
at the time just before a reaction
can be  deduced from its preimage and
the neighbors of the preimage according
to the following criterion:
If, $\tau$ time earlier, among the preimage cell and its
neighbors there were both types
$A$ and $B$ present,
then the cell must have become C during the
last reaction; if
all of them were of one type only (apart from
C, which is inert), the
cell inherits the type of its preimage.
This means that we  unify the advection-reaction
process in one mapping connecting the cell contents just before reactions.
In all experiments the reaction range
$\sigma$ is on the order of
the lattice size $\varepsilon_0$.



\section{Results}

\subsection{Autocatalytic reaction: $A+B \rightarrow 2 B$}

\begin{figure}[htbp] 
\centering{\epsfig{file=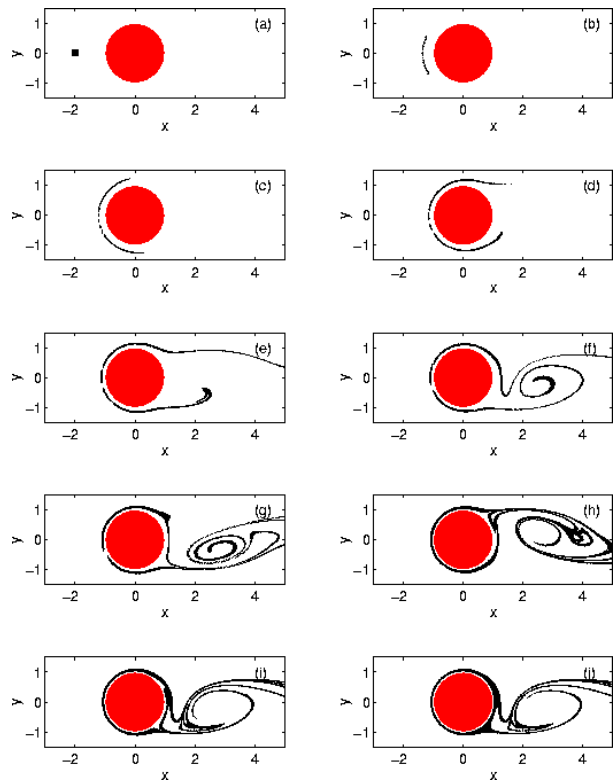,width=0.95\hsize}}
\vspace*{0.5cm}
\caption{Time evolution of a  seed of $400$ $B$  particles (black)
placed in the flow in front of the cylinder
(on a square of linear size $0.2$) at time $0$.
As the auto-catalytic
reaction evolves having $A$ as the background particles (white),
the amount of $B$ increases and traces out a complicated object 
in the wake of the cylinder. After some initial increase,
a steady state sets in.
The snapshots (a-j) are taken at times
$t=0,0.2,0.4,0.6,0.8,1.0,1.2,1.6,2.0$, and $3.0$
respectively,
right before a reaction takes place. The computational domain
$-3 < x < 5$ and $-1.5 < y < 1.5$  covers both the incoming and
mixing regions.
The time lag between reactions is $\tau=0.2$,
consequently $\nu=0.072$, and
the lattice size is $\varepsilon_0=\sigma=0.01$}\label{time-ev}
\end{figure}

Initially,
we introduce a seed of reagent $B$  in front
of the cylinder.
Since there are only two species in the
system, we monitor only reagent
$B$. Values referring
to  material $A$ inside
the computational domain
can be obtained from mass (in our
two-dimensional model, area)
conservation.
Figure~\ref{time-ev} displays  the
spreading of reagent $B$ (black)
in the course of time.
Note the rapid increase of the $B$ area and the quick
formation
of a filamental structure that becomes
stationary after a few time units, but changes periodically with the
period of the flow.

\begin{figure}[htbp] 
\centering{\epsfig{file=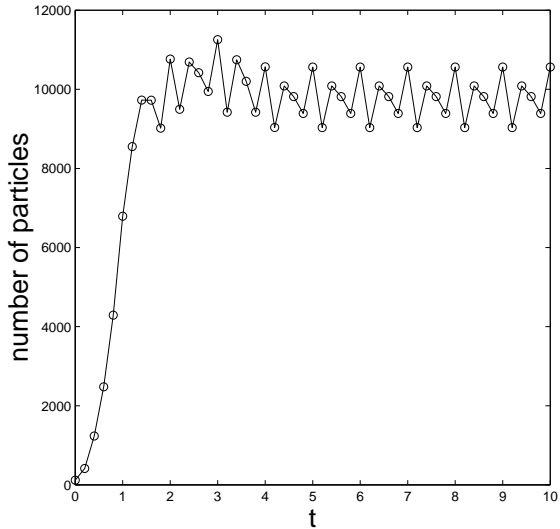,width=0.95\hsize}}
\vspace*{0.2cm}
\caption{The dependence of the number 
$ {\cal A}_B^{(n)}(\tau)/\varepsilon_0^2$ ($\varepsilon_0=\sigma$)
of $B$ particles 
in the computational domain of Fig.~2 on time $t=n \tau$ right before
the reaction events. 
Note the stationary time-periodic behavior
reached after about four time units.}\label{time-dep}
\end{figure}

To support this qualitative observation,
Fig.~\ref{time-dep} shows the number of $B$ particles
in the computational domain as a
function of time. After four
periods,
a self-repeating time dependence sets in.

This means that the chemical reaction
takes over  the
flow's basic periodicity
and  reaches  a stationary state:
the number of
cells being born in the reaction is the same
as the number of  cells
 escaping due to the advection dynamics.
 
\begin{figure}[htbp] 
\centering{\epsfig{file=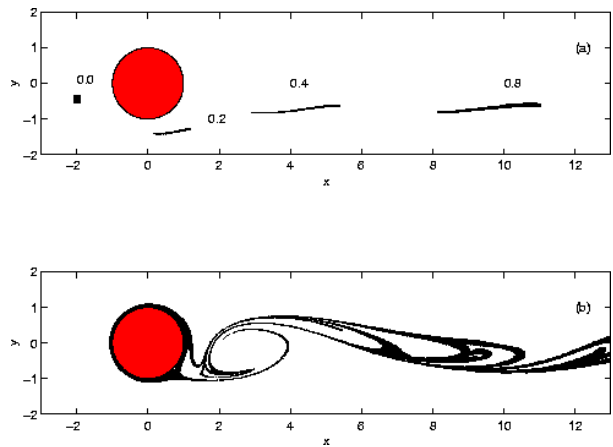,width=0.95\hsize}}
\vspace*{0.2cm}
\caption{(a) Time evolution of a droplet of the same size as
used in Fig.~\ref{time-ev}(a) but placed off axis for $\tau=0.2$
and lattice size $\epsilon_0 = \sigma = 0.01$.
The distribution of B particles is
shown at instants $t=0,0.2,0.4$~and~$0.8$.
No B particle is  in the wake of the cylinder after a time $1.0$.
(b) The same as Fig.~\ref{time-ev}(j) just in a more elongated  frame.
Note the downstream widening of the filaments.}\label{F:offax}
\end{figure}

In fact, owing to a special symmetry, which is not present
in the case of general obstacles,
the flow is
reflection symmetric with respect to the $x$ axis
after a time shift of one-half. Therefore,
the product distribution is of period $1/2$ \cite{rem1}.

\begin{figure}[htbp] 
\centering{\epsfig{file=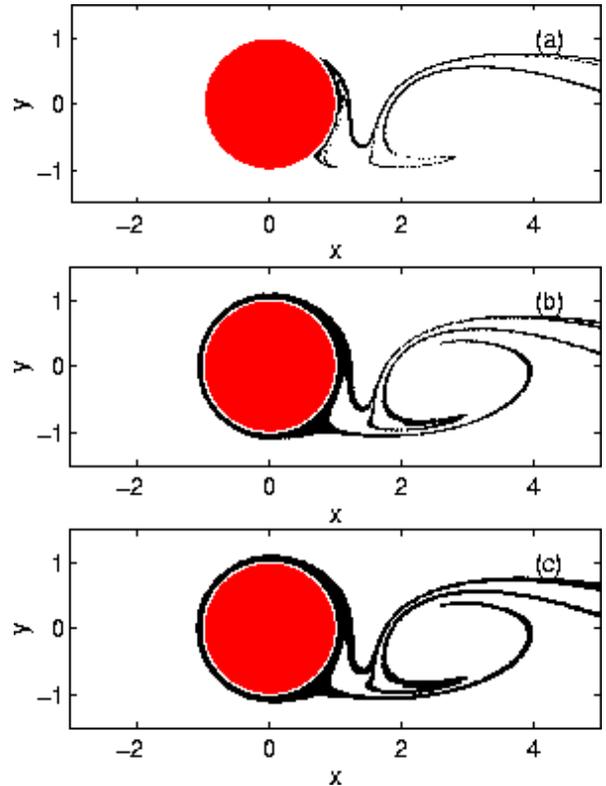,width=0.95\hsize}}
\vspace*{0.2cm}
\caption{(a) The unstable manifold
of the chaotic saddle in the reaction free
flow generated by distributing $20.000$ 
passive point particles (black dots) on  
short segments along the local
unstable direction of three basic fixed points of 
the Lagrangean dynamics, and iterating them forward in time 
over several periods.
(b) and (c) shows the $B$ particle distribution at
$t=10.0$ just before and just after the reaction, respectively.
Note the accumulation of $B$ along the 
unstable manifold and note its sudden broadening
from (b) to (c). The parameters are as in Fig~2.}\label{unstable}
\end{figure}

The outcome of the dynamics depends
strongly on the initial position
of the seed particle. If the initial
B droplet is off axis
[as in Fig.~\ref{F:offax}(a)], it does not
penetrate the mixing region
in the wake of the cylinder, and the initial droplet
is just simply stretched 
before the whole amount of B
is washed downstream. 
One can observe that the size of the compact patch B
increases due to the autocatalytic process
as time goes on.
Note that in this
case no material B remains in the
mixing region and the reaction
dynamics dies out in
any fixed observation region
of finite size.
This behavior is to be compared with cases
where the droplet penetrates the mixing region. 
To sharpen the contrast, in Fig.~\ref{F:offax}(b) we
display the $B$ distribution of Fig.~\ref{time-ev}(j)
in a much longer region downstream. This clearly indicates that material
$B$ is now present at {\em any} instant of time at {\em any}
$x$-value in the wake. The gradual broadening of the stripes of product
downstream is due to the autocatalytic feature of the process 
(and would not be present in the case of collisional reactions).

In what follows we  focus on such nontrivial
cases in which the droplet penetrates the mixing region. 
To understand the dynamics of Fig.~\ref{time-ev},
we recall that the tracer dynamics is governed
by a chaotic saddle in the wake
 of the cylinder. Passive tracers
coming close to the chaotic saddle
spend a long time in the mixing region
before being advected away along
the unstable manifold of the chaotic saddle
[cf. Fig.~\ref{unstable}(a)].
Thus tracers having spent  long time in the mixing
region   accumulate on the unstable manifold.
A comparison of Figs.~\ref{time-ev}(i-j)
and  \ref{unstable}(a)
provides numerical evidence for the accumulation of
material $B$
in stripes of finite widths along this manifold.

In order to gain more insight
into the reaction dynamics,
Figs.~\ref{unstable}(b) and
(c) show the reagent
distribution just before and
just after the autocatalytic reaction takes place,
respectively, in the steady state.
In the first case, the $B$ distribution has a rather
scanty appearance, while
right after the autocatalytic
reaction  most of the filaments
of the manifold are washed
 out due to a sudden widening.
The two pictures correspond to
two different coverages of the
fractal manifold. Just before the reaction,
the unstable manifold is covered
with stripes of average width
$\varepsilon^*(\tau)$,
while just after the reaction
with  $\varepsilon^*(0) > \varepsilon^*(\tau)$.
The sudden increase of the
coverage width at certain times
 is due to our modeling of the
chemical reaction as a ``kicked'' process.
In the case of time continuous reaction obtained in
the limit $\tau \rightarrow 0$
this feature is not present, but
the fact that material $B$ occupies a fattened-up
fractal remains unaltered.
We do not want to restrict our
attention to the time continuous case,
because the more general setting
might also be  of relevance to
biological phenomena, as discussed in the Introduction.

\begin{figure}[htbp] 
\centering{\epsfig{file=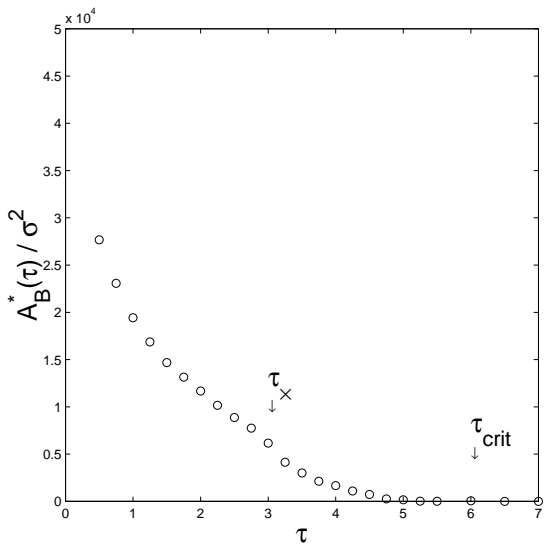,width=0.95\hsize}}
\vspace*{0.2cm}
\caption{Dependence of the number ${\cal A}_B^*(\tau)/\sigma^2$
of $B$ particles just before reaction in the steady state
on the time lag $\tau$ (all other parameters are as in
Fig.~2, with the exception of the reaction range
which is $\sigma=\varepsilon_0=0.003$, and of  that a boundary layer of width
$0.01$ has been cut out around the cylinder surface).
Note the sudden decrease around $\tau_{\times } \approx 3$
($\nu_{\times }\approx1.1$), leading to
a complete disappearance of $B$ from the mixing region, 
indicating an emptying transition,
around $\tau_{crit}=6$.}
\label{tau-dep}
\end{figure}

One of the most interesting quantities to follow is the
change of  the number of $B$ particles with the time lag $\tau$ (or $\nu$)
in the steady state, as shown in Fig.~\ref{tau-dep}.
Observe
the monotonic decrease and observe that for relatively large $\tau$
values ($\tau > \tau_{crit}$) no particle remains in the wake.
This
indicates the existence of an
``emptying transition''.
For reactions taking place rather
seldomly, the effect of the
advection by the background flow
is so strong  that no $B$ particles survive
the time lag $\tau$ in the wake
of the cylinder, and therefore, no
reaction  takes place in the mixing region.
Such emptying transition occurs if $\tau$ is on the order of 
 the chaotic
lifetime, and hence $\nu_{crit}\approx 1$.

\subsection{Collisional reaction: $A+B \rightarrow 2 C$}

\begin{figure}[htbp] 
\centering{\epsfig{file=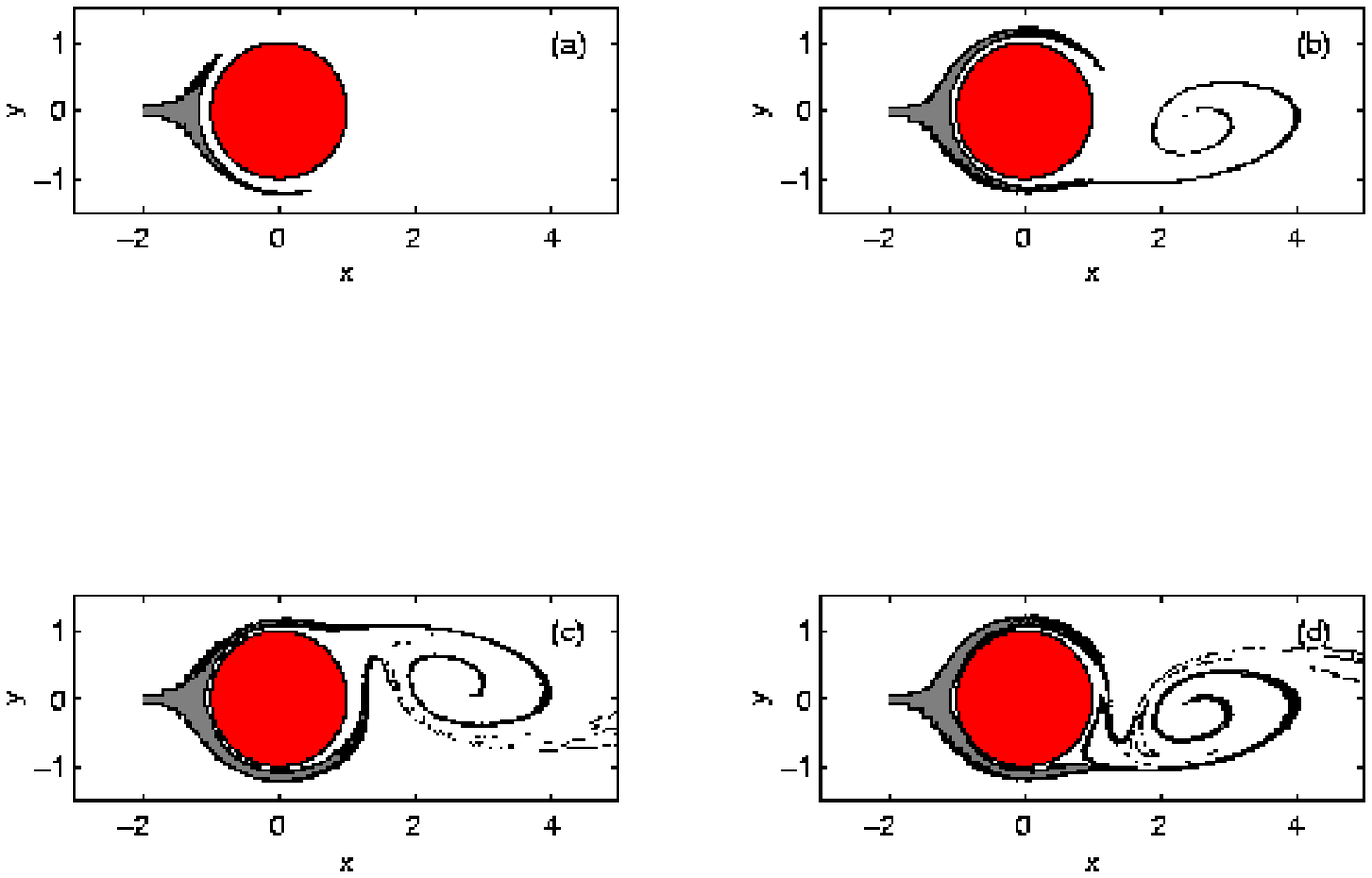,width=0.95\hsize}}
\vspace*{0.2cm}
\caption{
Time evolution for the collisional reaction $A+B \rightarrow 2C$.
Snapshots (a-d) of the surroundings of the cylinder are taken at times
$0.5,1,1.5$ and $2$, respectively,
right before the reactions. The region $-3<x<5$, $-1.5<y<1.5$
is shown. 
Initially the flow consists of material
$A$, into which
we introduce reagent $B$ along the line segment
$x=-2$, $-0.05<y<0.05$ ($l=0.1$).
White area denotes the background material $A$, grey is reagent $B$,
while black is the product $C$ separating the former constituents.
The
reaction range is $\sigma =0.011$, $\varepsilon_0=0.01$, 
while the time lag is $\tau =0.05$.
These imply an injection rate of $14.000$ particles
per unit time.}\label{fig:snap}
\end{figure}

Initially the flow consists of material
$A$, into which
we inject reagent
$B$ continuously, along a line segment
of length $l$ perpendicular to the background flow
in front of the cylinder.
As time evolves, material $C$ is produced.
Figure~\ref{fig:snap} displays 
typical snapshots of the surroundings
of the cylinder. The narrow stripes of constituent
$C$ (black)  separating
the areas occupied by background
material $A$ (white) and reagent $B$ (grey) are clearly visible.
Both  $B$ and $C$ stripes are
now pulled along the unstable
manifold of the chaotic saddle
forming lobes behind the cylinder.
This implies again that the reaction
essentially takes place along this manifold
[cf. Fig.~\ref{unstable}(a)].

\begin{figure}[htbp] 
\centering{\epsfig{file=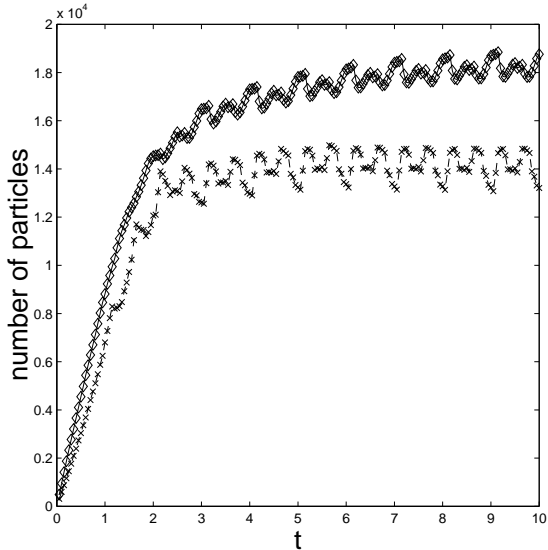,width=0.95\hsize}}
\vspace*{0.2cm}
\caption{
Time evolution of the number of cells occupied by $B$
(shown by diamonds) and $C$ (crosses) right
before the reaction in the numerical experiment of Fig.~7.
Note the asymptotic stationarity reached by both components.
}\label{fig:MCT}
\end{figure}

Figure~\ref{fig:MCT} shows the
time evolution of the number
of cells occupied by $B$ and $C$.
After a short transient, a saturation is reached.
This means, that the number of
$C$ cells born in the reaction is the same
as the number of $C$ cells
 escaping due to the emptying  dynamics.

As in the autocatalytic
reaction, the approached stationary value of $C$
depends on the reaction range
$\sigma$ and the reaction time $\nu$.
The number of $B$ cells also
converges to a stationary value after
sufficiently long time that
depends not only on
$\sigma$ and $\nu$, but also
on the velocity $v_0$ of the background flow,
and on the width $l$
since the amount of inflowing $B$  per unit time
is $v_0 l$.
The number of $B$ cells increases, in spite of the reaction
that consumes $B$, due to the flow that brings initially
more and more $B$ particles into the mixing region.
A periodic pulsation can
be observed after saturation has set in,
just like in the autocatalytic case.
By increasing the time lag $\tau$, we observe that
material $B$ is not arranged along a fractal set in the wake,
or does not reach the wake at all,
due to the finite resolution of the grid. This already happens
 for not too large values of the
grid size $\varepsilon_0$.
Because we are interested
in the pronounced fractal structures
formed, we keep the dimensionless reaction time considerable below unity
in order to ensure that activity extends into the wake
along the unstable manifold.


\section{Theory}

\subsection{Autocatalytic reaction: $A+B \rightarrow 2 B$}

\noindent
{\bf Basic dynamics}

\bigskip

Let ${\cal A}_{B}(t)\equiv{\cal A}_{B}^{(n)}(s)$ denote the area
occupied by 
reagent $B$.
Here $s\in \left[0,\tau\right]$ is the
time after the $n$th reaction.
Thus the total physical time is $t=n\tau+s$.
During the time interval of length $\tau$
only the dynamics of the chaotic flow controls the motion of the two
components. In a fixed region of observation, say in a rectangle in the
wake, ${\cal A}_{B}^{(n)}(s)$ decreases
according to the escape rate $\kappa$ of the chaotic saddle as
\begin{equation}
\frac{d{\cal A}_{B}^{(n)}}{ds}=-\kappa {\cal A}_{B}^{(n)}, ~~~0\le s<\tau,
\label{decay}
\end{equation}
provided the material is distributed along sufficiently
narrow stripes around the unstable manifold at $s=0$.
Solving this equation we
find that the area occupied by material $B$ at the end of the interval
$\tau$ is
\begin{equation}
{\cal A}_{B}^{(n)}(\tau)={\cal A}_{B}^{(n)}(0) \; e^{-\nu}
\label{eq:dynauto}
\end{equation}
with $\nu=\kappa \tau$ \cite{rem2}.

\begin{figure}[htbp] 
\centering{\epsfig{file=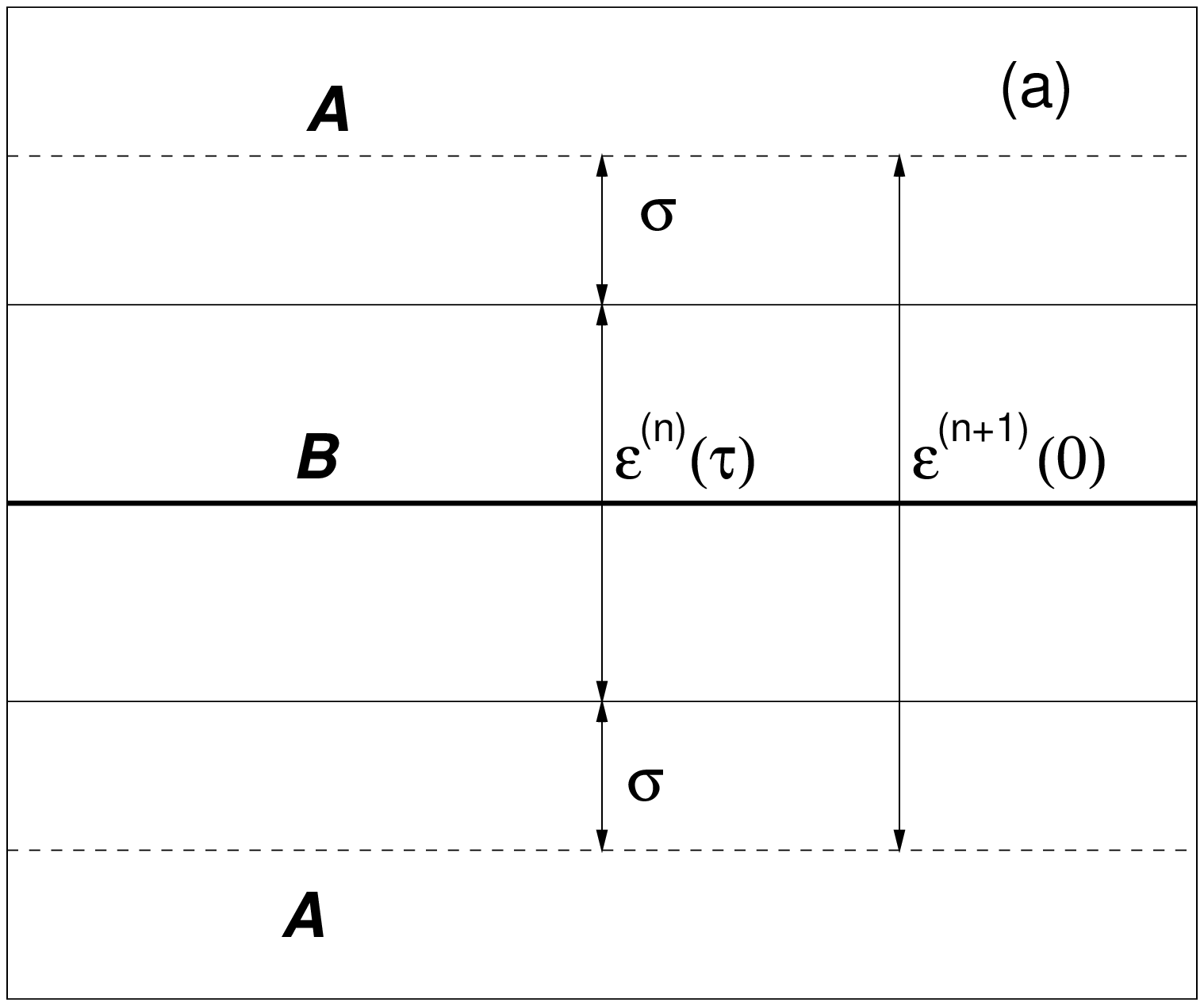,width=0.65\hsize}}\\
\vspace*{1.0cm}
\centering{\epsfig{file=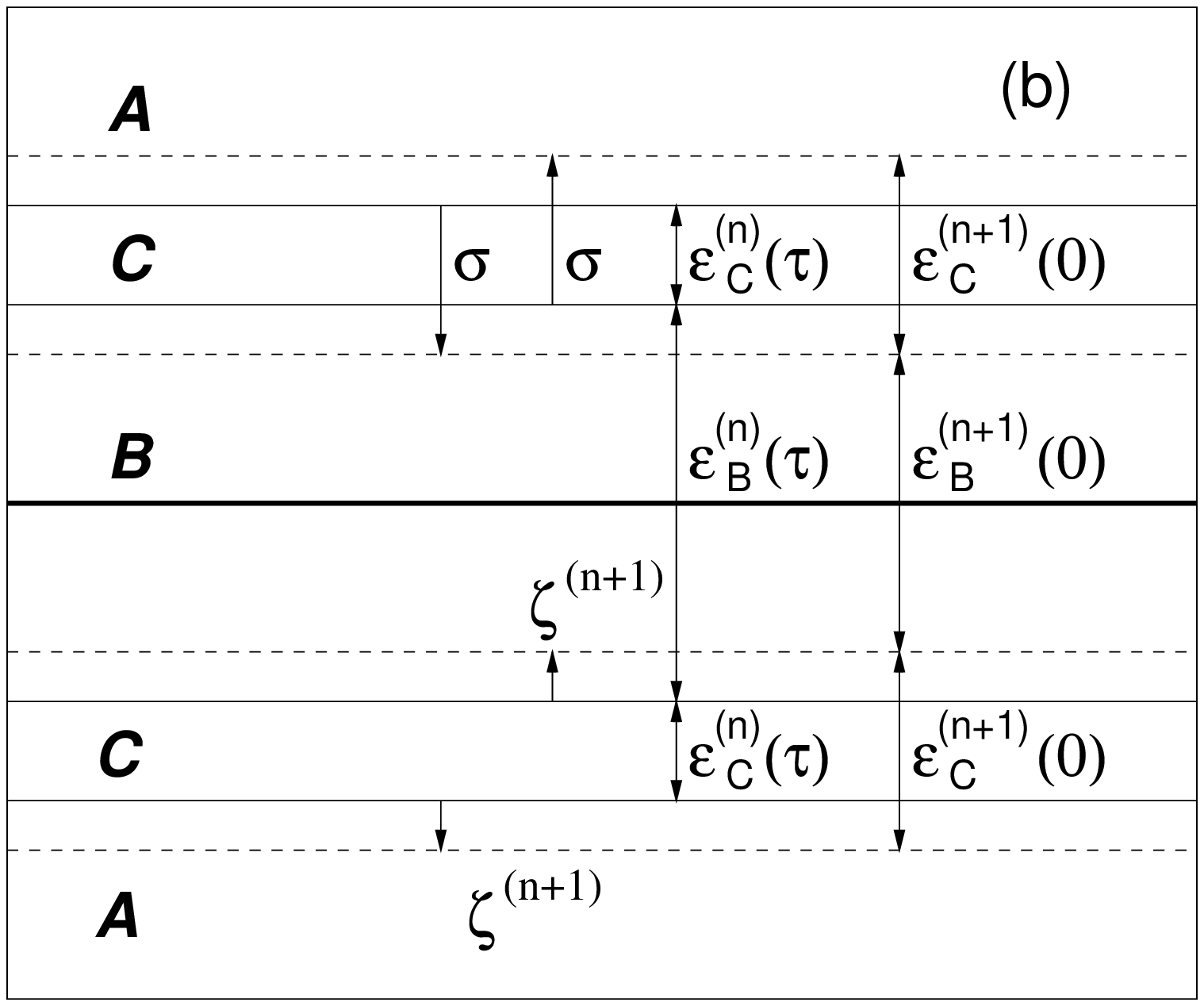,width=0.65\hsize}}
\vspace*{0.6cm}
\caption{
Schematic diagram of the components' distribution
along the unstable manifold (bold line)
before and after the $n$th reaction.
The borderline between the components is denoted by solid (dashed)
lines before (after) the reaction. (a) Auto-catalytic reaction.
(b) Collisional reaction.
Here all reagents $A$ and
$B$ within a distance $\sigma$ are transformed  into $C$ 
in the course of reaction.
The broadening of the half width of each $C$ stripe 
at reaction $n+1$ is denoted by $\zeta^{(n+1)} \ge 0$.
}\label{fig:theo}
\end{figure}

In order to determine the new area
${\cal A}_{B}^{(n+1)}(0)$ right after
the $(n+1)$th reaction [which is expected to
be larger than
${\cal A}_{B}^{(n)}(\tau)$],
we recall that
after sufficiently long time from the onset of the reaction
[when the exponential decay law (\ref{eq:dynauto}) holds]
the area of $B$ is pulled into narrow 
stripes of more or less
{\em constant} widths along the unstable manifold [cf. Fig.~\ref{fig:theo}(a)].
In other words, the fractal unstable manifold is {\em fattened up}
with material $B$ which has a {\em nonzero} area.
Let $\varepsilon(t) \equiv \varepsilon^{(n)}(s) \ll 1$ denote the {\em average}
width of the stripes at
time $t=n\tau+s$.
This means that on scales longer than or equal to $\varepsilon^{(n)}(s)$
the territory occupied by $B$ appears to be a fractal of the same
dimension $D_0$ as the unstable manifold. Covering the full area
occupied by $B$
at any time instant
by squares of linear size $\varepsilon\ge\varepsilon^{(n)}(s)$,
the number of boxes needed for
this coverage behaves as $N(\varepsilon)={\cal H}\varepsilon^{-D_0}$.
Here ${\cal H}$ is a constant
characterizing
the geometry, the so-called {\em Hausdorff measure} (or area) of the
manifold. It can be obtained by
determining the (unstable) manifold of the advection dynamics
of the
reaction-free flow. In what follows we assume ${\cal H}$ to be known.

It is worth noting that although the dimension is independent on the
instant at which the snapshot is taken, the Hausdorff measure is not.
Since the flow is periodic,  $\cal H$ is periodic with the period of the flow:
${\cal H}(t)={\cal H}(t+1)$.
(In fact, due
the special reflection symmetry explained in Sec.~III.A,
$\cal H$ is periodic with period $1/2$.)
For convenience, we choose the period of the flow
to be a multiple
or a divisor of the time lag: $M \tau =1$, where $M$ or $1/M$ is an
integer, respectively.
For time lags shorter than the flow's period, the
period contains an integer number of reactions, otherwise
the time lag is an integer multiple of the period.
Thus ${\cal H}^{(n)}(\tau)
= {\cal H}^{(n+1)}(0)$
$\equiv {\cal H}^{(n)}$ is 
$M$ or $1/M$ periodic as a function of $n$.
Since
$\varepsilon^{(n)}(s)$ is the smallest box size with which the
fractality of the reagent can be felt, the area
${\cal A}_{B}^{(n)}(s)$
of
$B$ can be written at any time as
\begin{equation}
{\cal A}_{B}^{(n)}(s)={\cal H}^{(n)}(s) \;
 {\varepsilon^{(n)}(s)}^{2-D_0}.
\label{eq:map01}
\end{equation}

To determine the dynamics of the covering width
$\varepsilon^{(n)}(s)$ we observe that, at
any reaction, there is a change  due
to the sudden increase of the product area [see Fig.~9(a)].
Since the filaments are
locally smooth {\em lines},
this widening is
{\em proportional} to the reaction range
 $\sigma$. We can thus write
\begin{equation}
{\varepsilon^{(n+1)}(0)}=
{\varepsilon^{(n)}(\tau)}+\sigma  d^{(n)}.
\label{eq:eps}
\end{equation}
If the widening were exactly orthogonal
 to the manifold,
and all the stripes occupied by $B$
 were nonoverlapping,
the coefficient $d^{(n)}$ would be
 $2$. After some time, however,
initially distinct stripes start to overlap.
This leads to a  change in the effective free surface
available for the reaction.  The phenomenological
factor $d^{(n)}$ introduced
in Eq.~(\ref{eq:eps}) describes both the
effect of the geometrical
shapes (stripes being tilted)
and the effect of overlap, on the average.
It depends on the time instant $n$,
and on the flow
parameters. 
The time evolution of the shape
factor $d^{(n)}$ is unknown {\em a priori}.
We shall see, however, that the
 theory becomes consistent
with the numerical observations if,
after a long time, the shape factor
takes over the period of the flow. Thus we assume that
$d^{(n)}$
is also $M$ or $1/M$ periodic in $n$.

By taking into account Eqs.~(\ref{eq:dynauto}-\ref{eq:map01}),
the area ${\cal A}_{B}^{(n+1)}(\tau)$ before the $(n+1)$st reaction 
can be written as
\begin{eqnarray}
{\cal A}_{B}^{(n+1)}(\tau) & = &{\cal H}^{(n+1)}
\left[ {\varepsilon^{(n+1)}
(\tau)} \right]^{2-D_0} \nonumber \\
& = &e^{-\nu}
{\cal H}^{(n)}
\left[ {\varepsilon^{(n+1)}(0)} \right]^{2-D_0}. \nonumber
\end{eqnarray}
{From} this we find that
 the average widths $\varepsilon^{(n+1)}(\tau)$ just before reaction
 $(n+1)$ is proportional to $\varepsilon^{(n+1)}(0)$, the width right after
 the $n$th reaction: 
$${\varepsilon^{(n+1)}(\tau)}=\varepsilon^{(n+1)}(0) \left[
\frac{e^{-\nu} {\cal H}^{(n)}}{{\cal H}^{(n+1)}} \right]^{1/(2-D_0)}.$$ 
Then from
(\ref{eq:eps}), 
a closed recursion relation follows for the 
widths  just before the
reactions:
\begin{equation}
{\varepsilon^{(n+1)}(\tau)}=\left[
\frac{e^{-\nu}{\cal H}^{(n)}}{{\cal H}^{(n+1)}}
\right]^{1/(2-D_0)}
\;\left[
{\varepsilon^{(n)}(\tau)}+\sigma d^{(n)} \right].
\label{eq:mapeps}
\end{equation}
In view of Eq.~(\ref{eq:map01}), this implies a recursion  for the $B$ area
as
\begin{equation}
{\cal A}_{B}^{(n+1)}(\tau)= e^{-\nu}
  \left\{ \left[{\cal A}_{B}^{(n)}(\tau)\right]^{1/(2-D_0)} +
  g^{(n)}  \sigma \right\}^{2-D_0}.
  \label{eq:mapauto}
\end{equation}
This {\em reaction equation}
is a discrete dynamics for ${\cal A}^{(n)}_B(\tau)$
expressing the amount of $B$ before a reaction in terms
of the amount before the previous reaction.
Since the quantities $d^{(n)}$ and ${\cal H}^{(n)}$ 
appear in a specific combination for this map,
we have introduced the shorthand notation
\begin{equation}
g^{(n)}\equiv d^{(n)}
\left.{\cal H}^{(n)}\right.^{1/(2-D_0)}.
\label{d}
\end{equation}
In contrast to the width dynamics (\ref{eq:mapeps}),
the area dynamics
contains all geometrical contributions via $g^{(n)}$, which
is called therefore the geometrical factor.

Equation (\ref{eq:mapauto}) belongs to the same class of dynamics
as recursion relations of the type $x_{n+1}=f(x_n)$,
like e.g.\ the famous logistic map \cite{Ott}. It is
one-dimensional and strongly dissipative,
therefore it describes the convergence towards asymptotic motions
which are represented by  attractors.
If ${\cal H}^{(n)} \equiv {\cal H}$ and $g^{(n)}\equiv g$ is constant,
a fixed point  of the system is found from
${\cal A}_{B}^{*}(\tau)= {\cal A}_{B}^{(n+1)}(\tau)=
{\cal A}_{B}^{(n)}(\tau)$
in the form of
\begin{equation}
{\cal A}_{B}^{*}(\tau) \equiv {\cal H}{\varepsilon^*}^{(2-D_0)} =
\left( \frac{\sigma g}{e^{\nu/(2-D_0)}-1}\right)^{2-D_0}.
                          \label{eq:MB0}
\end{equation}
This is the
area occupied by reagent $B$
right before a chemical reaction takes
place in the stationary state. The
area of $B$ right after the
reaction is
a factor $e^{\nu}$ larger.

\begin{figure}[htbp] 
\hspace*{0.72cm}\centering{\epsfig{file=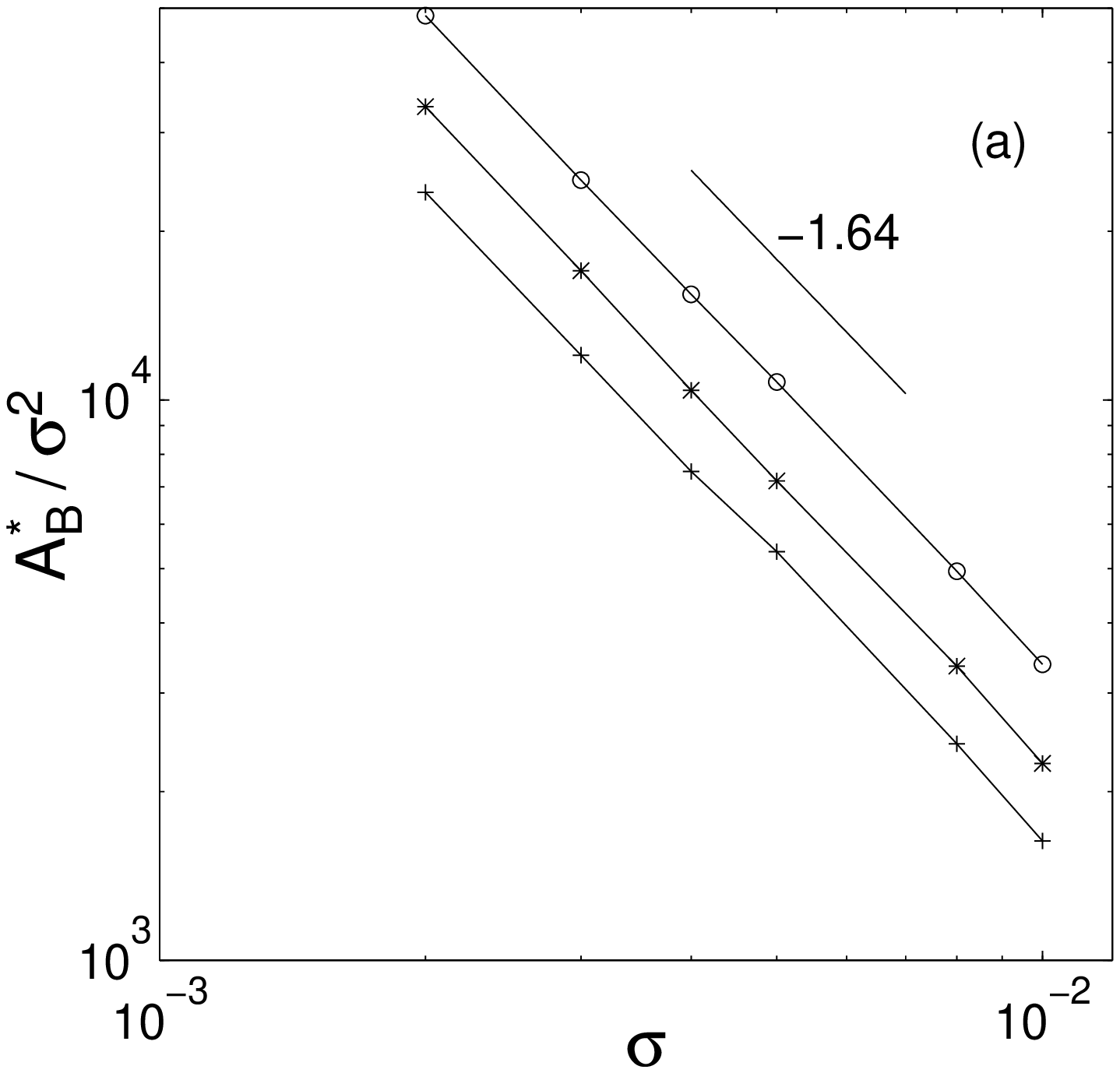,width=0.70\hsize}}\\
\centering{\epsfig{file=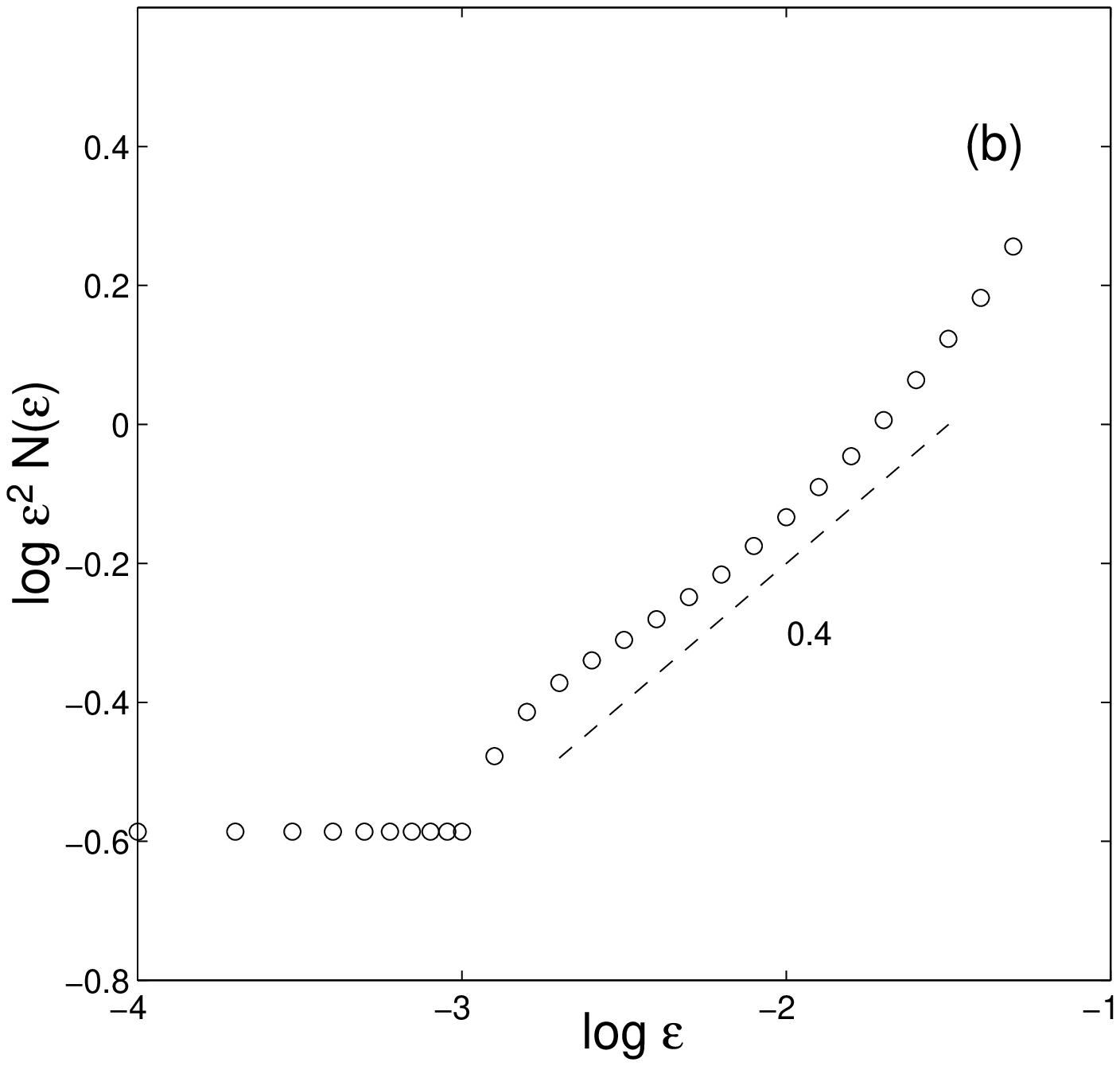,width=0.55\hsize}}\\
\hspace*{1.5cm}\centering{\epsfig{file=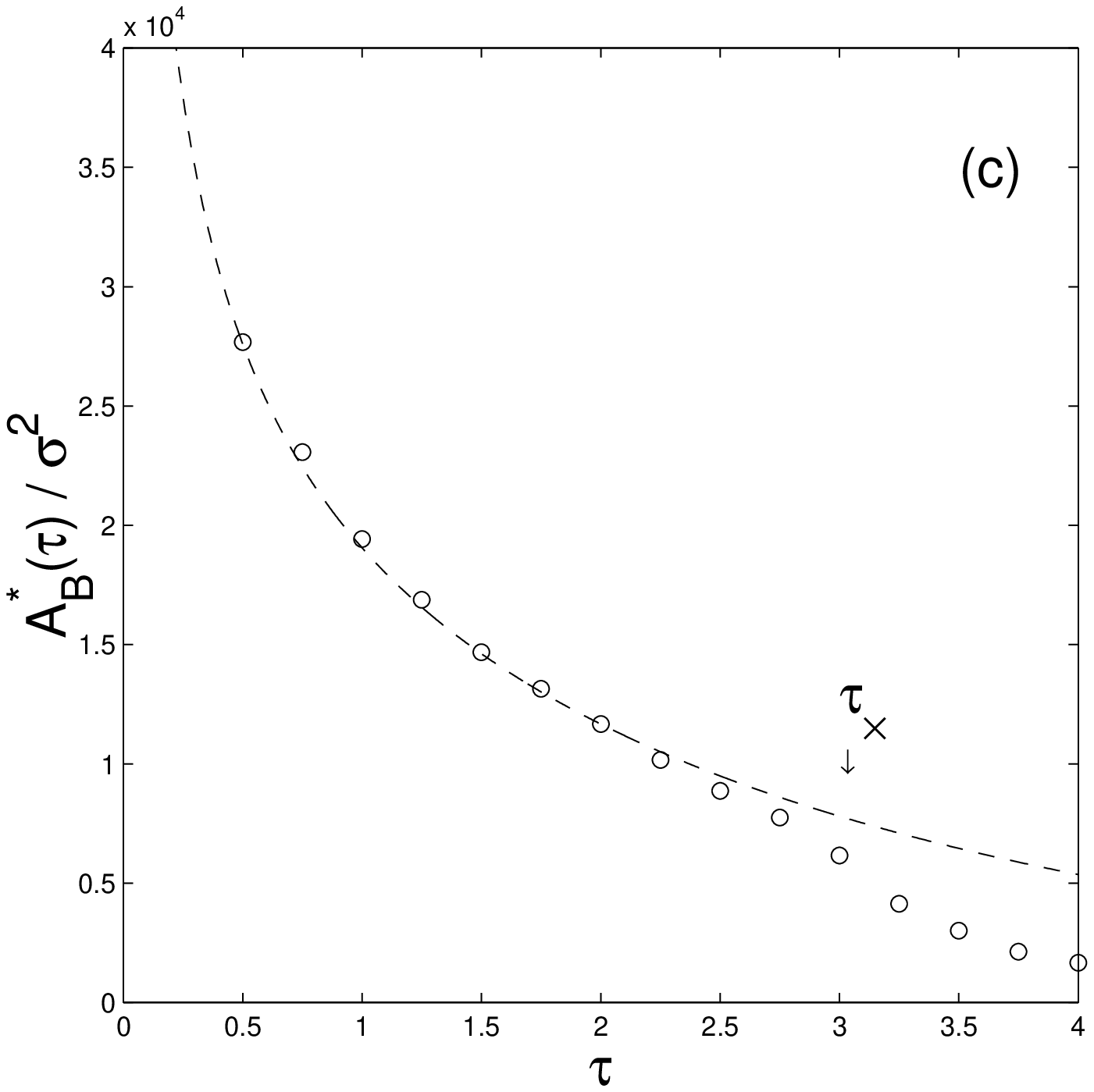,width=0.70\hsize}}
\vspace*{0.2cm}
\caption{
Scaling properties in the autocatalytic reaction.
(a) The area ${\cal A}_B^* \equiv {\cal A}_B^*(\tau)$ 
occupied by  $B$ in the steady state
 scales as ${\cal A}_B^* \sim \sigma^{2-D_0}$
 at different values of the
time lag $\tau$ ('o' $\tau = 1$, '*' $\tau=2$, '+' $\tau = 3$)
with $D_0=1.64$, 
in good agreement with the fractal dimension
of the unstable manifold of the reaction free flow.
Several runs were carried out with
different values of the reaction range
$\sigma$, which is set equal to the lattice constant
$\sigma=\varepsilon_0$.
(b) Number $N(\varepsilon)$ of boxes of size $\epsilon$, covering the
$B$ material distribution right before the reaction,
shows a fractal
scaling.
On the $\log_{10}{\varepsilon^2 N(\varepsilon)}$
vs. $\log_{10}{\varepsilon}$ plot an exponent
$0.4$ appears implying  $D_0=1.6$.
The crossover value is
$\epsilon^* \approx 0.003$.
Below this scale the
fractal structure is completely washed out and a trivial exponent $0$
emerges. The parameters are $\sigma = 0.001$, $\tau = 1$.
(c) The small to medium time lag range of Fig.~6, ($0 < \tau < 3$)
fitted to
(\ref{eq:MB0}) with $g=5.5$, $\kappa=0.36$, $D_0=1.61$, and
$\sigma = 0.003$.}
\label{fig:scaling1}
\end{figure}

In the more general case when $d^{(n)}$,
${\cal H}^{(n)}$, and thus $g^{(n)}$, are periodic with some
integer $M$, a limit
cycle of 
period $M$ is the attractor. The active process becomes thus
synchronized to the underlying flow dynamics.
Due to the linearity of Eq.~(\ref{eq:mapeps}), and the
fact that the factor $e^{- \nu/(2-D_0)}<1$,
the limit cycle attractor  of (\ref{eq:mapauto}) is always stable (in contrast
to the logistic map).

Equation~(\ref{eq:MB0}) can be used to
 determine approximately the mean value of
the geometrical factor $g^{(n)}$  and its oscillation amplitude about
the mean in a temporally periodic steady state. For the case
of Fig.~3, when the attractor is of period-5, we obtain
$g=16 \pm 3$.
Note that
for integer values of $\tau \ge 1$ (when $1/M$ is integer, too), $d^{(n)}$
and ${\cal H}^{(n)}$ are
constant since  the periodicity of the flow is unity. Then
(\ref{eq:MB0}) holds again exactly.

We have carried out a series of numerical experiments for such cases to  
carefully check the validity of
(\ref{eq:MB0}).
Fig.~\ref{fig:scaling1}(a)
shows the scaling of
$\ln{{\cal A}_B^*(\tau)}$
vs.\ $\ln{\sigma}$.
A scaling exponent of 0.36
emerges that is close
to the theoretically predicted value of $2-D_0\approx0.39$.

A quantitative measure of the fractality of the $B$
pattern can be obtained by
computing the ``box-counting''
dimension. The number of boxes of linear size $\epsilon$
that contain material
$B$ is expected to scale as $\epsilon^{-D_0}$.
Fig.~\ref{fig:scaling1}(b) shows a fractal scaling with
$1.6$ for $\epsilon > \epsilon^*$
and a sudden crossover
at $\epsilon^*$ due to the fattening-up of the fractal
at small scales. This $\epsilon^*$
is the average width of the
coverage of the unstable manifold and is found to be
$\epsilon^*=0.003 \pm 0.0002$. A comparison with
(\ref{eq:MB0}) yields that  the shape factor is on the order of $d=0.6$.

Fig.~\ref{fig:scaling1}(c)  shows a fit of
Eq.~(\ref{eq:MB0}) to the measured values
of ${\cal A}_{B}^{*}(\tau)$.
A good quality fit has been obtained with
the geometrical  factor
$g=5.5$ up to moderate values of the time lag $\tau$.

Observe from formula (\ref{eq:MB0}) that, in the
limit as $\tau \to \infty$, the {
area} of  material $B$ vanishes,
${\cal A}_{B}^{*}(\tau) \to 0$. This does not necessarily
mean that there is no material $B$ left in the mixing
region. In fact, there is {\em always} some material
$B$ left in this region, in the form of a fractal set of zero area. The
$\tau \to \infty$ limit corresponds to the case where
the $B$ set literally coincides with the unstable
manifold of the chaotic set. Since during a reaction
the $B$ set fattens up with  average width
proportional to $\sigma$, the fattened-up $B$ set should
not have vanishing area as $\tau \to \infty$. In fact,
assuming that 
$g^{(n)}$ is
constant, and using
Eqs.~(\ref{eq:dynauto}) and (\ref{eq:MB0}), the limit
$\tau \to \infty$ gives:
\begin{equation}
\lim_{\tau \to \infty} {\cal A}_{B}^{*}(0) =
\lim_{\tau \to \infty} e^{\nu} {\cal A}_{B}^{*}(\tau) =
(\sigma g)^{2-D_0} \neq 0. \label{nemnulla}
\end{equation}
Thus, taking the limit $\tau \to \infty$ in (\ref{eq:MB0})
is only consistent with the chemical framework, i.e., with
a continuous distribution of particles.
Of course, the numerics
can deal  with finite grid size only. However, when the
grid size is small enough, we get 
a good agreement with our formulas, for not
too large time lags, as shown in Fig.~\ref{fig:scaling1}(c).

\bigskip

\noindent
{\bf Emptying transition}

\bigskip

When deriving the theory, we do not make use
of the finiteness of the grid, which
can also be interpreted
as the finiteness of the particle size.
It is this finite size effect which leads to
the {\em emptying transition}
for large time lags.
If during the advection
dynamics of duration $\tau$ the
number of escaping particles exceeds the number
of new $B$ particles created at the next
reaction, then the balance favors the total
extinction of material $B$ from the wake.
This happens at a {\em critical} value for
the dimensionless reaction-time, $\nu_{crit}$ .

We derive an expression for  $\nu_{crit}$
via defining the {\em productivity}
of the reaction.
The productivity of the chemical
reaction in the steady state
can be characterized by the ratio of
newly-born to parent particles $B$ as
\begin{equation}
S=[{\cal A}_B^*(0)-{\cal A}_B^*(\tau)]
/{\cal A}_B^*(\tau)=e^{\nu}-1. \label{gyerek}
\end{equation}
The productivity $S$, however,
cannot be arbitrarily large. An absolute
maximum $S_{max}$ exists,
since the number of cells inside
the reaction range $\sigma$ is {\em limited}.
This implies that
\begin{equation}
e^{\nu}-1\le S_{max}.   \label{eq:emptr}
\end{equation}
For $\nu$ smaller than
\begin{equation}
\nu_{crit}= \ln{(1+S_{max})}  \label{eq:TCR1}
\end{equation}
the productivity of the chemical reaction
grows exponentially with 
increasing  reaction time $\tau$ or $\nu$.
For $\nu > \nu_{crit}$, however,
$S$ does not grow further.
Using Eq.~(\ref{gyerek}) and the fact that
 $S\le S_{max}$, we have
\begin{equation}
{\cal A}_B^{*}(0)\le (1+S_{max})\;{\cal A}_B^{*}(\tau).
\end{equation}
Inserting expression (\ref{eq:dynauto}), this leads to
\begin{equation}
{\cal A}_B^{*}(0)\le (1+S_{max})\;
{\cal A}_B^{*}(0)\; e^{-\nu}.
\label{eq:TCR2}
\end{equation}
For $\nu > \nu_{crit}$ the quantity
$(1+S_{max}) e^{-\nu}$ is
less than $1$, therefore (\ref{eq:TCR2}) can only
hold for ${\cal A}_B^{*}(0)=0$.
 Thus, for $\nu > \nu_{crit}$,
the area of reagent $B$ quickly
drops to zero in time, and remains
only the background material $A$ in
the system in the steady state.
In our case $S_{max}=8$, thus the critical
reaction time lag becomes $\tau_{crit}=
2 \ln{3}/\kappa = 6.10$. Indeed, this
is confirmed by the measurements exhibited
in Fig.~\ref{tau-dep}.

The definition of the productivity leads
us to another important time
value, the {\em crossover time lag},
$\nu_{\times }$.
It represents the threshold
time lag of productivity for which the
effects of finite particle size (and grid)
come into play. Below this value, the filaments
are `tightly' covered by the $B$ cells, so on average
one particle is responsible for {\em at most two}
newborns, $S=2$.
At $\nu_{\times }$, the coverage of the filaments
is just about to break up, thus:
\begin{equation}
\nu_{\times } = \ln{(1+S)}=\ln{3}\;,
\end{equation}
which leads to $\tau_{\times } = \ln{3}/ \kappa =
\tau_{crit} / 2 = 3.05$. For values of $\tau$ larger than $\tau_{\times}$,
the theory described in the previous paragraph
breaks down since it does not make use of the
finite particle size. Indeed, a deviation from (\ref{eq:MB0})
shows up in Fig.~\ref{tau-dep} by $ \tau=3$.

\bigskip

\noindent
{\bf The continuous time limit}

\bigskip

Before taking the continuous time limit
$\tau \rightarrow 0$,
it is worth rewriting the reaction
equation (\ref{eq:mapauto})
in a different form.
If the coverage of the manifold by $B$ is relatively
wide, i.e.,
$\varepsilon^{(n)}(\tau) \gg g^{(n)} \sigma$,
we can     expand the
right hand side of (\ref{eq:mapeps})
to first order
to obtain
\begin{eqnarray}
{\cal A}_{B}^{(n+1)}(\tau) & = &
{\cal A}_{B}^{(n)}(\tau)\;
 e^{-\nu} \nonumber \\
& + &
\sigma g^{(n)}
\; (2-D_0)\;  e^{-\nu}
\left[ {\cal A}_{B}^{(n)}(\tau)
\right]^{-\beta}, \label{eq:mapautoold}
\end{eqnarray}
where
\begin{equation}
\beta =\frac{D_0-1}{2-D_0} \label{beta}
\end{equation}
is a nonnegative number (it takes on zero only
in nonchaotic flows).
This equation is equivalent to saying that the reaction
gain is proportional to the perimeter $\cal L$ of the
$B$ area times the reaction range
$\sigma$ \cite{TKPTG}, and that this gain can be
expressed to be proportional
to a power $-\beta$ of the area.
[With the identification $c \equiv g^{(n)} (2-D_0)$,
Eq.~(\ref{eq:mapautoold}) coincides with Eq.~(3) of Ref.\cite{TKPTG}.]
 Because 
exponent $\beta$ is
{\em negative}, the equation contains a
singular expression of
the $B$ area.
The fixed point (\ref{eq:MB0}) for
$\varepsilon^*$ is, in general,
of the order of $g \sigma$, and the
condition leading to this form
is typically not valid after a long time. Therefore,
Eq.~(\ref{eq:mapautoold}) can only hold for a transient period
before coming close to the attractor. If, however,
the time lag is small enough and decreases, $\nu \ll 1$, the region of
validity of (\ref{eq:mapautoold}) is increasingly longer.
It is thus not surprising that in the
continuous time limit,
the differential equation
obtained for the $B$ area is the
analog of the {\em singular} map (\ref{eq:mapautoold}):
\begin{equation}
\dot{{\cal A}}_B=-\kappa {\cal A}_B+g(t)
\; (2-D_0) v_r
  \left(
{\cal A}_B
\right)
^{-\beta}. \label{diffeq}
\end{equation}
Here ${\cal A}_B(t)$ denotes ${\cal A}_B^{(t/\tau)}(0)$
 or ${\cal A}_B^{(t/\tau)}(\tau)$
for $\tau \rightarrow 0$, and
\begin{equation}
v_r = \lim_{\tau \rightarrow 0} \frac{\sigma}{\tau}
\end{equation}
represents the { velocity} of the reaction front.

A comparison with the case of a uniform flow
[Eq.~(\ref{diffeqa})] shows that
Eq.~(\ref{diffeq})
represents a novel form of reaction equation containing
a {\em negative} power of the material 
content due to the fractality of the
unstable manifold.
Its special case obtained
for $D_0=1$ describes a surface reaction in the presence of a single
isolated hyperbolic orbit in the wake
\begin{equation}
\dot{{\cal A}}_B=-\kappa {\cal A}_B+g(t) \; v_r.
   \label{diffeq1}
\end{equation}
Note that this is already structurally similar to the
traditional equation (\ref{diffeqa}). But even for a single
periodic orbit the
decay is slower than without its existence
because the time scale set by the escape rate (which is
now just the Lyapunov exponent of the orbit) is typically much
longer than the characteristic time of the flow.

Altogether, Eq.~(\ref{diffeq}) {\em deviates in both terms} from
the traditional reaction dynamics (\ref{diffeqa}). {\em The decay is
slower while productivity is much faster than
in an unstructured flow.} Both effects are due to the
presence of  a chaotic saddle which produces an escape rate
$\kappa$ smaller than $v_0$ and a dimension $D_0$ bigger than one.

Our findings concerning the new form of the reaction equation
are similar in spirit to those
of Muzzio and Ottino \cite{Muzzio}.
They considered the effect of filamentation on chemical reactions
in closed containers without finding an explicit form 
for the reaction equations. 
The difference with us is due to 
the fact that we are studying now open flows with fractal patterns
of $D_0<2$, and our exponent $\beta$ is therefore  unique.

The
area occupied by $B$ in a stationary state
[$g(t)=g$]
follows from
$\dot{{\cal A}}_B=0$ as
\begin{equation}
{\cal A}_B^{*} \equiv {\cal H}{\varepsilon^*}^{2-D_0}=
  \left( \frac{ g \;(2-D_0)\;v_r}{\kappa}\right)^{2-D_0},
\label{cond}
\end{equation}
in accordance with the $\tau \rightarrow 0$ limit of
Eq.~(\ref{eq:MB0}). 
Equation~(\ref{cond}) has an important
 consequence for the velocity
of a reaction with visible fractal
properties. The latter can only be
seen if $\varepsilon^*$ is much
less than unity (the cylinder radius)
which implies, for $g$ factors of
the order of one, that
$v_r \ll \kappa < v_0$. Fractal product
distributions can only be expected
for reactions which are {\em slow}
 on the lifetime of chaos.
In this continuous time limit the
emptying transition does not occur,
since $e^{\nu}\rightarrow 1$, and
 therefore (\ref{eq:emptr}) is fulfilled.

It is worth mentioning briefly that
different continuous time limits
are also possible depending on the
assumptions  made on the model
parameters. One option is to split
the reaction range $\sigma$ into
a `deterministic'  part $\bar{\sigma}$ and a fluctuating
$\delta \sigma$, i.e.,
$\sigma= \bar{\sigma}+\delta \sigma$.
The reaction front velocity is
then defined via $\bar{\sigma}$, and
$\delta \sigma/\tau$ goes over as
a noise term $\xi$ in the continuous time limit. So the reaction
equation becomes a stochastic differential equation
\begin{equation}
\dot{{\cal A}}_B=-\kappa {\cal A}_B+
g(t)\; (2-D_0)\; (v_r + \xi)
  \left( {\cal A}_B \right)^{-\beta} \label{diffeqr}
\end{equation}
Note that the noise term $\xi$ appears in
this Langevin equation in
a {\em multiplicative} form. It's effect is
thus also influenced by the
fractality of the underlying manifold.


\subsection{Collisional reaction: $A+B \rightarrow 2 C$}

\noindent
{\bf Basic dynamics}

\bigskip
The area occupied by the materials $B$
and $C$ at time $s$ after the
$n$th chemical reaction is denoted by
${\cal A}_B^{(n)}(s)$ and
${\cal A}_C^{(n)}(s)$, respectively.
During the time interval
$\tau$ there is
no reaction, thus only the chaotic
advection and the injection of $B$
governs the dynamics of the area of $B$ and $C$ as
\begin{eqnarray}
\frac{d{\cal A}_B^{(n)}}{ds} & = &
v_0 l-\kappa {\cal A}_B^{(n)},
  \nonumber  \\
\frac{d{\cal A}_C^{(n)}}{ds} & =
& -\kappa {\cal A}_C^{(n)}.
\end{eqnarray}
Here $v_0$ denotes the velocity of
the inflow far upstream from the
cylinder, and
$l$ is the width of the stripe along which material
$B$ is being injected in the inflow region. The rest of
the inflow consists of $A$ only.
After solving these equations we find
the areas occupied by $B$ and
$C$ right before the next reaction to be:
\begin{eqnarray}
{\cal A}_B^{(n)}(\tau) & = &
{\cal A}_B^{(n)}(0) e^{-\nu}+
   \frac{v_0 l}{\kappa}\;
   (1-e^{-\nu}),   \nonumber  \\
{\cal A}_C^{(n)}(\tau) & = &
{\cal A}_C^{(n)}(0) e^{-\nu}.
\label{eq:MBC}
\end{eqnarray}
The reaction $A+B \rightarrow 2C$ takes place if
the distance between $A$ and $B$
is {\em less} then $\sigma$.
Thus, the amount
${\cal A}_B^{(n+1)}(0)$
of material $B$ and
${\cal A}_C^{(n+1)}(0)$ of
$C$ right after the $(n+1)$th
reaction becomes smaller than
${\cal A}_B^{(n)}(\tau)$
and larger than
${\cal A}_C^{(n)}(\tau)$, respectively.

The geometry is now somewhat more
involved than in the autocatalytic case.
The branches of
the unstable manifold are covered with
material $B$ in stripes of average
 widths $\varepsilon_B^{(n)}(s)$.
Adjacent to this, there are {\em two}
stripes of equal widths $\varepsilon_C^{(n)}(s) $
containing only material $C$  [Fig.~\ref{fig:theo}(b)],
while material $A$ is outside.
Thus, {\em both} materials $B$
and $C$ lie along a fattened-up
copy of the unstable manifold.
Note, however, that the amount of
fattening-up is different for these materials.

In an analogous way as for the autocatalytic reaction,
we introduce again a phenomenological
shape factor $d^{(n)}$.
Due to the effects of overlaps and
non-straight geometry of the unstable manifold,
we assume that
a reaction  takes place at time $(n+1)\tau$ if
$\varepsilon_C^{(n)}(\tau)$ is
smaller than the range
$\sigma d^{(n)} $. Here $d^{(n)}$
 is again
dependent on the time instant,  being $M$ or $1/M$ periodic in $n$
after sufficiently long times.
The broadening  of the half width of each $C$ stripe
(which is the same as the
corresponding decrease of the $B$ stripes)
in the $(n+1)$th reaction is
\begin{equation}
\zeta^{(n+1)} \equiv
\sigma d^{(n)} -\varepsilon_C^{(n)}(\tau).    \label{eq:Ddyn}
\end{equation}
This leads to the following change of the
covering stripe widths:
\begin{equation}
\varepsilon_B^{(n+1)}(0)=\varepsilon_B^{(n)}(\tau)
- 2 \zeta^{(n+1)}, \label{eq:mape0}
\end{equation}
and
\begin{eqnarray}
\varepsilon_C^{(n+1)}(0) & = & \varepsilon_C^{(n)}(\tau) + 2 \zeta^{(n+1)} 
\nonumber \\
                  & = & 2 \sigma  d^{(n)}  - \varepsilon_C^{(n)}(\tau).
\label{eq:mapde0}
\end{eqnarray}
Note that
Eq.~(\ref{eq:mape0}) is meaningful
only if the difference on the right hand side
is non-negative. This depends on the relation
between the variables $l$ and $\sigma$. For a fixed
$l$, $\sigma$ can be increased independently, and
at a certain value, the increase in the width of the
$C$-stripes exceeds the width of the $B$ stripes
during a reaction. 
{From} now on, we shall take $\sigma$ small enough,
and assume that the situation in Fig.~\ref{fig:theo}(b)
is valid at all times during the process.

Since the fractal scaling still holds for the actual
width of $B$ or $C$, at an arbitrary time, the areas are
\begin{eqnarray}
{\cal A}_B^{(n)}(s) & = & {\cal H}^{(n)}(s)\;
\left[ \varepsilon_B^{(n)}(s)\right]^{2-D_0}, \nonumber \\
{\cal A}_C^{(n)}(s) & = & 2 {\cal H}^{(n)}(s)\;
\left[ \varepsilon_C^{(n)}(s)\right]^{2-D_0}.   \label{eq:Mdeps}
\end{eqnarray}
If the  first equation is also applied to the
inflow region, 
we have to assume that the width $l$ of the injection
of $B$ is on the same order 
as $\varepsilon_B^{(n)}(s)$,
so that its two-dimensional character cannot
yet be seen on this scale.
Substituting these
into (\ref{eq:MBC}), we find two relations
connecting  the widths taken at different
times:
\begin{eqnarray}
{\cal A}_{B}^{(n+1)}(\tau) & = & {\cal H}^{(n+1)}
\left[ {\varepsilon_B^{(n+1)}(\tau)} \right]^{2-D_0} \nonumber \\
& = & e^{-\nu}
{\cal H}^{(n)}
\left[ {\varepsilon_B^{(n+1)}(0)} \right]^{2-D_0}
+ \frac{v_0 l}{\kappa}  (1- e^{-\nu}), \nonumber 
\end{eqnarray}
and  
\begin{eqnarray}
{\cal A}_{C}^{(n+1)}(\tau) & = &2 {\cal H}^{(n+1)}
\left[ {\varepsilon_C^{(n+1)}(\tau)} \right]^{2-D_0} \nonumber \\
& = & 2 e^{-\nu} 
{\cal H}^{(n)}
\left[ {\varepsilon_C^{(n+1)}(0)} \right]^{2-D_0} \nonumber
\end{eqnarray}
with ${\cal H}^{(n)}\equiv
{\cal H}^{(n)}(\tau)=
{\cal H}^{(n+1)}(0)$.
In view of
(\ref{eq:mape0})
 and
(\ref{eq:mapde0})
we obtain a coupled set of recursions
for either the
stripe widths
or for
${\cal A}_B^{(n)}(\tau)$ and
${\cal A}_C^{(n)}(\tau)$. The dynamics
of the covering
stripe width
$\varepsilon_C$
for material $C$ has the simpler form:
\begin{equation}
{\varepsilon_C^{(n+1)}(\tau)}=
\left[
\frac{e^{-\nu} \;{\cal H}^{(n)}}{{\cal H}^{(n+1)}}
\right]^{1/(2-D_0)}
\;\left[
2 \sigma  d^{(n)}  -\varepsilon_C^{(n)}(\tau)  \right].
\label{eq:mapde}
\end{equation}
Consequently, the change  of the half width is:
\begin{eqnarray}
\zeta^{(n+2)} & = & -
\;\left[
\frac{e^{- \nu}{\cal H}^{(n)}}{{\cal H}^{(n+1)}}
\right]^{1/(2-D_0)}\;
\zeta^{(n+1)} \nonumber \\
& + &  \sigma \;\left\{ d^{(n+1)}-
d^{(n)}
\;\left[
\frac{e^{- \nu}{\cal H}^{(n)}}{{\cal H}^{(n+1)}}
\right]^{1/(2-D_0)}\;
\right\}.
\label{eq:mapd}
\end{eqnarray}
The recursion for the
stripe width
$\varepsilon_B$ covering reagent $B$
is then obtained from Eqs.~(\ref{eq:MBC}),
(\ref{eq:mape0}) and (\ref{eq:Mdeps}) as
\begin{eqnarray}
{\varepsilon_B^{(n+1)}(\tau)} & = &
 \left\{  \frac{v_0 l}
{\kappa {\cal H}^{(n+1)}}\; (1-e^{-\nu})
\right. \nonumber \\
+e^{- \nu}
\frac{{\cal H}^{(n)}}{{\cal H}^{(n+1)}}
 & & \left. \left[
{\varepsilon_B^{(n)}(\tau)}- 2
\zeta^{(n+1)}
\right]^{2-D_0} \right\}^{1/(2-D_0)}.\label{eq:mape}
\end{eqnarray}

Finally,
we find that the recursion relations for
the areas are given by the following reaction equations:
\begin{eqnarray}
 {\cal A}^{(n+1)}_B(\tau) & = &
      \frac{v_0 l}{\kappa} (1-e^{-\nu})
  + e^{-\nu}
      \left[ \left( {\cal A}^{(n)}_B(\tau)
      \right)^{1/(2-D_0)} \right.
      \nonumber \\
 & + & 2 \left.
      \left( \frac{{\cal A}^{(n)}_C(\tau)}{2}
      \right)^{1/(2-D_0)} -
      2 g^{(n)} \sigma
      \right]^{2-D_0},
      \nonumber  \\
\frac{{\cal A}^{(n+1)}_C(\tau)}{2} & = &
   e^{-\nu}
      \left[ 2 g^{(n)} \sigma - \left( \frac{{\cal A}^{(n)}_C(\tau)}{2}
               \right)^{1/
      (2-D_0)}
      \right]^{2-D_0}.
           \label{eq:MAPABCC}
\end{eqnarray}
Here $g^{(n)}$ denotes the same geometrical factor
(\ref{d}) as defined for the autocatalytic process.
This form implies that the area dynamics for both
$B$ and $C$ only depends parametrically on $g^{(n)}$,
while the $\varepsilon_B$ and $\varepsilon_C$  stripe dynamics
contain $d^{(n)}$ and ${\cal H}^{(n)}$ as
independent parameters.

Note that the $C$-reaction (and $\varepsilon_C$)
is decoupled from
$B$ (and $\varepsilon_B$),
and component $B$ simply follows
the $C$-dynamics:
the second of Eqs.~(\ref{eq:MAPABCC}) is independent of
the $B$ component, which may seem surprising at
first sight.
It was derived with the assumption that $\sigma$
is small enough, and that the average width
of material $B$ is large enough to furnish enough reagent for each
reaction event:  the $B$ stripe is not totally consumed
during an instantaneous reaction.
Thus the dynamics of the reaction product $C$ depends only on
the actual width of the $C$
stripes which separate
$A$  from $B$. Between two
consecutive reactions, the
average $C$ width decreases,
making possible a widening at the next
reaction, and so on. Thus, the
presence of $B$ is necessary for
producing $C$, but if the $B$-area is
wide enough, the $C$-reaction becomes
independent of $B$.

In the case when the Hausdorff
measure ${\cal H}$ and the shape factor $d$ (and geometrical factor
$g$)
are $n$-independent, we find
a stationary state.
The average width of the $C$
stripes can be given explicitly:
\begin{equation}
\varepsilon_C^{*}(\tau)  =  \varepsilon_C^{*}(0) \;
e^{-\nu/(2-D_0)}=
  \frac{2 \sigma d}{e^{\nu/(2-D_0)}+1}.
  \label{fixde}
\end{equation}
Consequently, from Eq.~(\ref{eq:mapd})
\begin{equation}
\zeta^{*}  =  \sigma  \;d\;
\tanh \frac{\nu}{2(2-D_0)}
\label{decsillag}
\end{equation}
follows.
For the average width of the $B$ stripes
an implicit equation is obtained
\begin{eqnarray}
{\varepsilon_B^{*}(\tau)}^{2-D_0}  =
  \frac{v_0 l}{\kappa {\cal H}}\;(1-e^{-\nu})
 & + & e^{- \nu}  \left[
{\varepsilon_B^{*}(\tau)}- 2 \zeta^{*}
 \right]^{2-D_0}.
\label{eq:width}
\end{eqnarray}
Since in this equation all terms are expected to be of the same order
of magnitude, we find that
\begin{equation}
\varepsilon_B^* \approx \left( \frac{v_0 l}{\kappa {\cal H}} \right)^{1/(2-D_0)}
\end{equation}
holds.
It means that the coverage width of the inflow of reagent $B$ is
of the order of the average coverage width in the mixing range,
thus our earlier assumption on the validity of
Eq.~(\ref{eq:Mdeps}) is fulfilled.
The fixed point of recursion
(\ref{eq:mapde}) and (\ref{eq:mapd}) is an attractor for any
parameter since $e^{-\nu/(2-D_0)}<1$.

The fixed point expressions for the
areas occupied by $B$ and $C$
are obtained as
${\cal A}_B^{*}(s)  =
    {\cal H}\varepsilon_B^*(s)^{2-D_0}$ and
${\cal A}_C^{*}(s)  =  2{\cal H}
    \varepsilon_C^{*}(s) ^{2-D_0}$,
respectively. Thus,
\begin{equation}
{\cal A}_{C}^{*}(\tau) \equiv 2 {\cal H}{\varepsilon_C^*}^{(2-D_0)} =
2 \left( \frac{2 \sigma g}{e^{\nu/(2-D_0)}+1}\right)^{2-D_0},
                          \label{eq:MC0}
\end{equation}
and
\begin{eqnarray}
& \;\;& \left[
 {\cal A}_B^{*}(\tau) e^{\nu}
+\frac{v_0 l}{\kappa}(1-e^{\nu})
\right]^{1/(2-D_0)} =
  {{\cal A}_B^{*}(\tau)}^
{1/(2-D_0)}  \nonumber \\
 & - &
2 g \sigma +
2 \left(\frac{{\cal A}_C^{*}(\tau)}{2}\right)^{1/(2-D_0)}.
\label{eq:FIXB}
\end{eqnarray}
Note that although (\ref{eq:MC0}) is in some sense the analogue
of (\ref{eq:MB0}), the particular $\nu$-dependence is different:
we have now a Fermi-Dirac distribution, while
(\ref{eq:MB0}) was of Bose-Einstein type. This difference is due
to the fact that the increase of $\varepsilon_C$ in a reaction step
is not a constant, rather it is  proportional to $\zeta$ which also depends on
$\varepsilon_C$ itself [cf. (\ref
{eq:mapde0})].
{From} these two relations a $g$-independent form
follows
\begin{eqnarray}
& \;\;& \left[
 {\cal A}_B^{*}(\tau) e^{\nu}
+\frac{v_0 l}{\kappa}(1-e^{\nu})
\right]^{1/(2-D_0)} =
  {{\cal A}_B^{*}(\tau)}^
{1/(2-D_0)}  \nonumber \\
 & + &
\left(1- e^{\nu/(2-D_0)}\right)\;
\left(\frac{{\cal A}_C^{*}(\tau)}{2} \right)^{1/(2-D_0)},
\label{eq:FIXBC}
\end{eqnarray}
which does not contain any free parameters.

When the periodicity of ${\cal H}$, $d$ and
$g$ is pronounced,
recursions (\ref{eq:MAPABCC}) typically
 possess a limit cycle attractor
corresponding again to a full
synchronization to the flow
dynamics.

The consistency of expressions
(\ref{eq:MC0}),
(\ref{eq:FIXB}) and
(\ref{eq:FIXBC}) with the numerics
is verified
in different ways
(we fixed $v_0=14$ and  $l=0.1$).
The comparison of the results shown for a period 20 limit cycle
steady state in Fig.~8 with the
expression (\ref{eq:MC0}) yields $g=42 \pm 4$
for the geometrical factor.
This value is also
consistent with (\ref{eq:FIXB}) and the measured $B$ area.
In simulations using different grid sizes $\varepsilon_0$,
we find similar geometrical factor values that are slightly
increasing with $\varepsilon_0$.
We also evaluate the ratio of the left and right hand sides
of the fit-free relation (\ref{eq:FIXBC}) at different grid sizes
$\varepsilon_0=0.007$, $0.01$ and $0.013$ and
time lags $\tau=1/80, 1/40, 1/20$ and $1/10$. The ratio is
in all cases between 0.9 and 1.0 with an average
$0.96\pm 0.05$, $0.95\pm 0.04$ and $0.94\pm 0.03$
for the $\varepsilon_0$ values investigated, respectively.
In view of the fact that
$g^{(n)}$
is not a constant, since the time lag is definitely below
unity, the agreement, within an accuracy of 10 percent is
satisfactory because this is exactly
the amplitude of fluctuations in $g$.

\bigskip

\noindent
{\bf The continuous time limit}
\bigskip

\noindent

Before taking the time continuous limit $\tau \rightarrow 0$,
it is worth again rewriting the reaction equation (\ref{eq:MAPABCC})
in a different form.
If the coverage of the manifold by $B$ and $C$ is 
large with respect to the amount of broadening, i.e., if
$\varepsilon_B^{(n)}(\tau), \varepsilon_C^{(n)}(\tau) \gg \zeta^{(n+1)}$, we can
expand the
right hand sides of (\ref{eq:MAPABCC})
to first order in $\zeta$ using (\ref{eq:mapde}) and
(\ref{eq:mape}) to obtain
\bigskip
\begin{eqnarray}
 & &{\cal A}^{(n+1)}_B(\tau)  =  {\cal A}^{(n)}_B(\tau)e^{-\nu} +
      \frac{v_0 l}{\kappa}\; (1-e^{-\nu})  \nonumber \\
 & - & 2(2-D_0)e^{-\nu}\;
 \frac{\zeta^{(n+1)}}{d^{(n)}}
       g^{(n)}
      \left[ {\cal A}^{(n)}_B(\tau)
      \right]^{-\beta},
       \nonumber  \\
 & & \frac{{\cal A}^{(n+1)}_C(\tau)}{2}  =
     \frac{{\cal A}_C^{(n)}(\tau)}{2} e^{-\nu} \nonumber \\ 
 & + &
     2  (2-D_0) e^{-\nu}\;
 \frac{\zeta^{(n+1)}}{d^{(n)}} g^{(n)}
       \left[\frac{{\cal A}^{(n)}_C(\tau)}{2}
       \right]^{-\beta}.   \label{eq:MAPABCCold} 
\end{eqnarray}
Here $\beta$ is the same exponent (\ref{beta}) as in the autocatalytic
reaction.
The fixed point (\ref{fixde}) for $\varepsilon_C^*$ is, in general,
on the order of $\zeta^*$ and the condition leading to this form
is typically not valid after a long time. Therefore,
Eq.~(\ref{eq:MAPABCCold}) can only hold for a transient period
before coming close to the attractor. If, however,
the time lag is small enough and decreases, $\nu \ll 1$, the region of
validity of (\ref{eq:MAPABCCold}) is increasingly longer.

Next we take
the continuous time limit.
Let $f(t)$ denote the $\tau \rightarrow 0$ limit of any function
which takes the form
$f^{(n)}(s)$ at time $t=n \tau + s$ in the discrete time
representation, with $t$ kept fixed.
Let us first consider the recursion (\ref{eq:mapd})
of $\zeta$ and divide it by the shape factor $d$.
The basic observation is that due to the minus sign
in the first term on the right hand side, the recursion
in the limit $\tau \rightarrow 0$
does not go over into a differential equation
for $\zeta/d$.
Rather it shows that this ratio goes to zero linearly with the time lag.
Therefore it is natural to define a
reaction front velocity as
\begin{equation}
v_r(t) = \lim_{\tau \rightarrow 0}
 \frac{\zeta^{(n+1)}}{\tau d^{(n)}}.
\label{vr}
\end{equation}
It should be emphasized that this property means that
the broadening of the $C$ stripes should tend to zero
in the continuous time limit. Therefore, at any time
$\varepsilon_C(t)=\sigma d(t)$ holds.
Thus, in contrast to the autocatalytic reaction,
the reaction range needs not to go to zero now, since another
quantity, the actual broadening has to. In fact,
$\sigma$ is finite, and is a measure of the  front velocity.
A substitution of this form into  (\ref{eq:mapd}) yields the
explicit expression
\begin{equation}
v_r =  \frac{\sigma}{2} \left[
 \frac{\kappa }{(2-D_0)}+ \frac{d}{dt} \ln{g(t)} \right].
\end{equation}
Since $g$ is periodic with the flow's periodicity,
the front velocity is also periodic.

{From} recursion (\ref{eq:MAPABCCold})
we obtain for the differential equations of the $B$ and $C$ areas
in the limit
$\tau \rightarrow 0$
\begin{eqnarray}
\dot{{\cal A}}_B & = & -\kappa
{\cal A}_B+v_0 l-q(t)
  \left( {\cal A}_B
  \right)^{-\beta},      \nonumber  \\
\dot{{\cal A}}_C/2 & = & -\kappa {\cal A}_C/2 +q(t)
  \left( {\cal A}_C/2
  \right)^{-\beta},
  \label{eq:contABC}
\end{eqnarray}
where the coupling constant in both equations is the same
$q(t)= $ $2(2-D_0)$
$v_r(t) g(t)$. The two dynamics become entirely decoupled in this limit.
Notice the singular terms on the right hand sides again.

The fixed point value of the $C$ area [for $d(t) \equiv d$,
${\cal H}(t) \equiv \cal H$, $q(t) \equiv q$] is
\begin{equation}
{\cal A}_C^{*}  = 2 (q / \kappa)^{2-D_0}
\end{equation}
The $B$ area has also a fixed point solution,  but it is implicit:
\begin{equation}
{\cal A}_B^{*}  =
{\cal H}{\varepsilon_B^*}^{2-D_0} =\frac{v_0 l}{\kappa}
- \frac{q}{\kappa}
\left[{\cal A}_B^{*}
\right]^{-\beta}.
\label{eq:ABcsillag}
\end{equation}
The front velocity 
\begin{equation}
v_r  =
\frac{\sigma \kappa}{2(2-D_0)} =\frac{q}{2(2-D_0) g}
\end{equation}
is now a constant entirely determined by the reaction free flow's
parameters and the reaction range. This implies $\varepsilon_C^*=d \sigma$ in
accordance with the $\nu \rightarrow 0$
limit of  equation  (\ref{fixde}).


\section{Discussion}

First we discuss a few aspects of the nonlinear dynamics which has not
been mentioned in the text.

The stable manifold  is a fractal curve leading particles
to the chaotic saddle. Points lying close to it reach  the saddle rapidly,
and their long lifetime
is due to a residence time in the neighborhood of the saddle. In other words,
initial conditions close to or far away from the stable manifold lead to
considerable lifetime  differences only after the saddle region
has been reached.
Thus, although the role of stable and unstable
manifolds seems to be symmetric in the problem, the residence 
about the stable manifold
is not long enough for an enhancement in activity. Therefore, no fattening-up
takes place along this manifold.
Anyhow, a nontrivial reaction can only occur if the
$B$ (and  $C$) initial conditions  intersect the stable manifold.
Otherwise, cases like the one shown  in Fig.~4(a) occur without
a fractal product distribution.

The advection dynamics is known to have  a
considerable nonhyperbolic component consisting of
points lying very close to the
cylinder surface. 
The nonhyperbolic part is characterized by a nonexponential
decay ($\kappa=0$)
and space-filling fractality ($D_0=2$) \cite{T0}.
Previous studies \cite{JTZa,JTZb} have shown, however,
that in the von K\'arm\'an flow model the resolution allowed in computer
simulations is still too crude to observe the nonhyperbolic effects
away from (but close to) the cylinder surface. The relative strength of the
hyperbolic component ensures that on the time and length scales used
we are able to work with a nontrivial fractal dimension
and a finite escape rate.
Apart from the nonhyperbolicity seen in  
the boundary layer around the obstacle, other
nonhyperbolic structures are expected to be seen on very small scales only.
The presence of a finite coverage with widths $\varepsilon^*$, however,
prevents us to reach these scales.
This is why the properties of the hyperbolic component play essential role
in the full process. Therefore, in any fixed frame in the
wake not overlapping with the boundary layer, the results of
the theory presented here are expected to hold \cite{rem3}.

Next we summarize those features of our model which are believed to
be general for active processes accompanied by weak
diffusion
in open flows with velocities faster than that of the reaction.

$\bullet$ Active processes take place about the unstable manifold
of the passive dynamics' invariant set. If the
dynamics is chaotic, the manifold is a fractal and, consequently,
the reaction leads to fractal patterns.

$\bullet$ Although the fractal itself is a set of measure zero,
the chemical products
are of {finite} amount due to the fattening up process.

$\bullet$ The fractal backbone results in an increase of the active
surface, it acts as a catalyst, and generates an enhancement in
activity as compared to flows with nonchaotic particle dynamics.

$\bullet$ The inclusion of weak molecular 
diffusion in the model can be regarded
as a random walk superimposed on the deterministic advection.
This would result in a further fattening-up of the fractal set, in
a renormalization of $\sigma$ and thus of $\varepsilon^*$.

$\bullet$ As a formal
consequence of the fractal backbone,
the product of the reaction obeys a singular scaling law.

$\bullet$ In spite of the passive tracers'
Hamiltonian dynamics, the active processes' equations are of
{dissipative} character possessing  attractors.

$\bullet$ Most typically,
a kind of {steady state} sets in after
sufficiently long times, a state which is synchronized with the flow's
temporal behavior.

$\bullet$ Fractality is independent on whether the traditional
reaction equations are linear or nonlinear since it is a
consequence of the advection dynamics' strong nonlinearity.

$\bullet$ The essential parameters for the chemical reaction  in the flow
depend on the parameters of the reactionless dynamics:
the escape rate $\kappa$ and fractal dimension $D_0$. These
in turn depend on parameters (like the Reynolds number)
of the underlying hydrodynamics.

$\bullet$ The derivation of the reaction equations
is similar to the derivation of macroscopic transport equations
from microscopic molecular dynamics. It seems that the presence of
ever refining fractal structures (which cannot be observed
directly with finite resolution) generate new terms in
the reaction equation leading to observable macroscopic
effects based on the fractal microstructures. They appear not only
in the averages but also in moments if a stochastic description is used.

All these features are expected to be present in realistic
chemically or biologically active environmental flows observed on
finite time scales.


\section*{Acknowledgements}

Useful discussions with
J. U. Grooss, P. Haynes, B. Legras,
H. Lustfeld, D. McKenna,
A. Mariotti, Z. Neufeld,  K. G. Szab\'o,
J. A. Yorke, and all the participants of
the ESF TAO Workshop `Chemical/Biological
Effects of Mixing', Cambridge, 1997 are acknowledged. \'AP and ZT thank
J. Kadtke, R.K.P. Zia and B. Schmittmann for their support
and  encouragement.
This research has been supported by the NSF through the Division of
Materials Research, by the DOE,
by the US-Hungarian Science and
Technology Joint Fund
under Project numbers 286 and 501, and by the
Hungarian Science Foundation T17493,
T19483.
One of us (G. K.) is indebted to the Hungarian-British Intergovernmental
Science and Technology Cooperation Program  GB-66/95
for financial support.


\newpage



\begin{thebibliography}{xx}
\bibitem{note} 
For the case  II.A  (II.B) 
a cell can have more than one preimages (images) due to the lattice
dynamics. This means that the number of the product particles is slightly
decreasing (increasing) in time
compared to one in the continuum description of the flow. 
The use of the different lattice dynamics 
is motivated by the observation that the  loss of $B$
generated by the method II.A  is compensated by
the strong autocatalytic spreading, while the same choice 
would have led to an underestimation of  product $C$ in the collisional
reaction.
\bibitem{rem1}
For general values of $\tau$, when
there is no reaction exactly at time $1/2$ or $1$,
the period of the fluctuations in the
number of $B$ particles
is  the smallest common multiple
of 1/2  and $\tau$.
\bibitem{rem2}
Due to the periodic motion of the unstable manifolds'
lobes, there might be a small oscillatory component present in  the
decay dynamics. The decay rate
$\kappa$ of (\ref{decay}) should be replaced by
$\kappa+f(t)$ with $f(t)$ as a periodic function
of small amplitude. In this case ${\cal A}_B^{(n)}(\tau)$ can
be expressed via an integral of the time-dependent decay rate.
\bibitem{rem3}
By not cutting out the boundary layer, we find that relation
(\ref{eq:MB0}) is still valid but with an effective escape rate smaller than
$\kappa=0.36$. This value, however, is dependent on  details,
like the time of observation or the position of the frame.
Interestingly, it is only the escape rate, whose value
is influenced considerably by the presence of the boundary layer.
Cutting out a layer of width $0.01$ does not modify  the fractal
dimension extracted from Figs.~\ref{fig:scaling1}(a),(b).
\end{thebibliography}
\end{document}